\documentclass[]{jpp}
\usepackage{graphicx}
\usepackage[utf8]{inputenc}
\usepackage[T1]{fontenc}
\usepackage{amsmath}
\usepackage{color}
\usepackage{mathtools}
\usepackage{bm}
\usepackage[mathlines]{lineno} 
\usepackage[]{natbib}
\usepackage{aas_macros}

\shorttitle{}
\shortauthor{Shalaby et al.}

\title[]{The growth of the longitudinal beam-plasma instability in the presence of an inhomogeneous background}

\author{
Mohamad Shalaby\aff{1}  \corresp{\email{mshalaby@live.ca}},
Avery E.~Broderick\aff{2,3},
Philip Chang\aff{4},
\\
Christoph Pfrommer\aff{1},
Ewald Puchwein\aff{1} \and 
Astrid Lamberts\aff{5}
}

\affiliation{
\aff{1} {Leibniz-Institut f{\"u}r Astrophysik Potsdam (AIP), An der Sternwarte 16, 14482 Potsdam, Germany}
\aff{2} {Department of Physics and Astronomy, University of Waterloo, 200 University Avenue West, Waterloo, ON, N2L 3G1, Canada}
\aff{3} {Perimeter Institute for Theoretical Physics, 31 Caroline Street North, Waterloo, ON, N2L 2Y5, Canada}
\aff{4} {Department of Physics, University of Wisconsin-Milwaukee, 3135 N. Maryland Ave., Milwaukee, WI 53211, USA}
\aff{5} { Universit\'e C\^ote d'Azur, Observatoire de la C\^ote
  d’Azur, CNRS, Laboratoire Lagrange, Laboratoire Art\'emis, Bd de l'Observatoire, CS 34229, 06304 Nice, cedex 4, France}
}

\begin{document}

\maketitle
\date{\today}
\begin{abstract}

We study the longitudinal stability of beam-plasma systems
in the presence of a density inhomogeneity in the background plasma.
Previous works  have focused  on the non-relativistic regime where
hydrodynamical models are used to evolve pre-existing Langmuir waves within inhomogeneous background plasmas.
Here, for the first time we study the problem with kinetic equations in a
fully-relativistic way.
We do not assume the existence of Langmuir waves, and
we focus on the rate and the mechanism by which waves are excited in such
systems from an initial perturbation.
We derive the structure of the unstable modes and compute an analytical approximation for their growth rates. 
Our computation is limited to dilute and cold beams, and shows an excellent agreement with particle-in-cell simulations performed using the SHARP code.
We show that, due to such an inhomogeneity, the virulent beam-plasma instabilities in the
intergalactic medium are not suppressed but their counterparts in  the solar wind can be suppressed as
evidenced by propagating type-III solar radio bursts.

\end{abstract}

\section{Introduction}

Dilute plasma beams propagating though ionized background
media are ubiquitous in astrophysical plasmas, which themselves span many
scales and parameters, e.g., beam-to-background density ratio and beam velocity.
Thus, understanding the stability of beam-plasma systems
is essential to modeling their evolution and understanding many
astrophysical phenomena.
Examples include AGN driven beam-plasma
instabilities in the intergalactic medium~\citep{blazari}, gamma-ray
bursts~\citep{GRB+BP-instability1,GRB+BP-instability2}, accretion disks around
black-holes~\citep{accretion1}, the solar wind~\citep{Ginzburg+1958}, pulsar wind~\citep{pulsar-winds+1978}, and
 relativistic jets from  AGNs~\citep{Jet-sims1,Jet-sims2}.
To study the stability of such systems, most analytical work has focused on the problem with a uniform background plasma number density.
These include studies using both hydrodynamical and more comprehensive kinetic descriptions of beam-plasma systems, see, e.g.,
\citet{bret2010}.

However, it is clear that in some astronomical contexts background inhomogeneity is a critical element.  For example, inhomogeneity is necessary to explain the apparent suppression of the non-relativistic plasma beams that are driven during type-III radio bursts
~\citep{Lin+1981}.
Estimates based on growth rates in
  the case of uniform  background plasmas imply fully thermalized beam-particles at 1
AU. 
In stark contrast, observations show that the beams persist and do not show the
expected plateau in their momentum distribution~\citep{Lin+1981}.
To explain this, hydrodynamical models of long-wave length ($\lambda \gg
\lambda_D $) and slowly varying Langmuir waves envelope, i.e., based on the high frequency limit of Zakharov equations~\citep{Zakharov-1972}, are developed.
These models assume the pre-existence
of Langmuir waves, i.e., assume that these are the unstable modes of the system
due to the beam-propagation; and investigate the evolution of
such wave packets in an inhomogeneous
medium~\citep{Ergun+2008,krafft-2013}.
These models provide a possible explanation of the observed wave clumping and apparent suppression of the beam instability in Type-III radio bursts.

On the other hand, one-dimensional models based on kinetic equations have been developed
\citep{Breizman+Ruytov_1970,breizman+1972,breizman+inhomo01,Nishikawa+1976}.
These models assume the validity of the uniform beam-plasma picture and study
how a non-uniform background number density changes the evolution and resonances
of the driven Langmuir waves, using the geometric-optic approximation.  While
these models succeed in explaining observations of type-III radio bursts, i.e.,
non-relativistic beam-plasma instabilities, 
they were used by~\citet{Miniati-Elyiv-2013} to imply an erroneous suppression of the instabilities in the
relativistic regime as shown by~\citet{sim_inho_18}.
Their PIC simulations show a
clear growth of the instabilities, very similar to the uniform case.

Here we  revisit both analytically and numerically the growth of longitudinal beam
plasma modes using the Vlasov-Poisson system.  We assume a quadratic
inhomogeneous structure in the background number density and derive the
fully-relativistic kinetic dispersion relation for this case.
We focus on the growth rate of the instability for dilute and 
cold beams from an initial perturbation, and derive the structure of 
the unstable modes for such system, i.e., we do not assume pre-existing 
Langmuir waves.
Various predictions, e.g., the rates of
wave growth and the shape of the unstable modes, are shown to have an excellent
agreement with PIC simulations.

This paper is organized as follows.  In Section~\ref{sec:formalism}, we present
the dispersion relation obtained from the linearization of the
Vlasov-Poisson equations in the presence of a quadratic  background density inhomogeneity.
Section~\ref{sec:NoBeam} drives the normal modes of such inhomogeneous systems
in the absence of beam-particles.  In Section~\ref{sec:WBeam}, we study the
effect of weak beams, i.e., the instabilities in presence of dilute and/or
relativistic cold-beams, on these normal modes by using analogies with first-order
perturbation theory.  In Section~\ref{sec:PICsims}, we present a list of
predictions from our computation and compare those to particle-in-cell (PIC)
simulations using the SHARP code~\citep{sharp}.  We discuss the implications of
this theory in inhomogeneous intergalactic and solar wind media in
Section~\ref{sec:application}, and summarize and conclude in
Section~\ref{sec:conclusion}.

\section{Formalism}
\label{sec:formalism}

It is often the case that dynamical time for large-scale structures substantially exceeds the relevant plasma timescales for beam-plasma instabilities.  This large separation in temporal scales admits a natural simplification of the problem: here, we focus on a beam-plasma system where electron-positron beams are propagating through a denser background of electrons and a fixed neutralizing protons. Two illustrative astrophysical applications, the intergalactic medium and solar wind, are presented in Section~\ref{sec:application}, where our assumption of a fixed-background approximation is demonstrated to be an excellent approximation.  Nevertheless, we expect this to have broad applicability to beam-plasma situations more generally.

We denote the phase space distribution functions of beam 
electrons/positrons by $f^{\pm}$ and for background electrons by $g$.
For such a case, the linearized
(first-order) Vlasov-Maxwell equations describe the evolution of longitudinal
modes, i.e., parallel to the beam direction; for detailed derivation, see, e.g.,
Section~4.2  of~\citet{ShalabyThesis2017}.  The resulting equations can be
re-written as an eigenvalue problem as follows:
\begin{align}
\label{eq:Eevol00}
&
\left[
k
+
\dfrac{ e^2}{m_e \epsilon_0 }
\int
du
\frac{\partial _u (f^{+}_0 + f^{-}_0) }{ \omega - k v }
\right]
E_1 (k,\omega)
+
\dfrac{e^2}{m_e \epsilon_0 }
\iint
dk^{'}
du 
\frac{ ~ \partial _u  \mathit{g}_0(k^{'},u)
}{ \omega - k v }
E_1(k-k^{'}, \omega)
=0
.
\end{align}
Here, $e$ and $m_e$ are the elementary charge and mass of electrons
respectively, $v$ is the velocity in phase space, $u = \gamma v$ 
is the spatial component of the four velocity with
Lorentz factor, $\gamma = 1/\sqrt{1-v^2/c^2}$,  $c$ is the speed of light,
$f^{\pm}_{0}$  are the
equilibrium phase space distribution function of
pair-beam plasma particles,  $\mathit{g}_{0}$ is the 
equilibrium phase
space distribution function of background electron plasma, and $E_1$
is the first order perturbation in the electric field.
The convolution in Equation~\eqref{eq:Eevol00} complicates finding solutions of this Equation.
However this can be greatly simplified when the inhomogeneity has a quadratic structure.
Therefore, in the following we consider an inhomogeneity such that the number density of the background electrons is
\begin{eqnarray}
n_g(x) &=& n_0 (1 + \epsilon x^2).
\end{eqnarray}
Assuming that $\mathit{g}_0(x,u) = n_g(x) g_0(u)$, as, e.g., in an isothermal plasma, we can write
\begin{eqnarray}
\mathit{g}_0(k^{'},u) = n_0 \left[ \delta(k^{'})  - \epsilon \delta^{''}(k^{'}) \right]  g_0(u),
\end{eqnarray}
where $\epsilon$  has dimensions of inverse length squared, and  we take $\epsilon \geq 0$, i.e., the inhomogeneity in the number density forms a quadratic bowl with a minimum at $x=0$.
In such a case, Equation~\eqref{eq:Eevol00} can be written as
\begin{align}
\label{eq:Eevol01}
&
\left[
k
+
\dfrac{ e^2}{m_e \epsilon_0 }
\int \frac{du }{ \omega - k v }
\partial _u (f^{+}_0 + f^{-}_0)
\right]
E_1 (k,\omega)
+
\left[
\omega_{0}^2
\int
du 
\frac{ ~ \partial _u g_0(u)
}{ \omega - k v }
\right]
(1 - \epsilon \partial^2_k )
E_1(k, \omega)
=0
,
\end{align}
where $\omega_{0}^2 =  e^2 n_0 / (m_e \epsilon_0) $ is the plasma frequency of the background electrons at $x=0$.

\section{Solution without beam}
\label{sec:NoBeam}

In this case, $f_0^{\pm}=0$, and Equation~\eqref{eq:Eevol01} becomes
\begin{align}
\label{eq:Eevol02}
&
\left[
k
+
\omega_{0}^2 
\int
du 
\frac{ ~ \partial _u g_0(u)
}{ \omega - k v }
\right]
E_1(k, \omega)
-
 \epsilon
\left[
\omega_{0}^2 
\int
du 
\frac{ ~ \partial _u g_0(u)
}{ \omega - k v }
\right] 
\partial^2_k
E_1(k, \omega)
=0
.
\end{align}
Because the equation is written in the frame of the background electrons, in which they only have thermal motions, the integral in Equation~\eqref{eq:Eevol02} can be solved in the non-relativistic limit~\citep{Boyd-Sand:03,linear-paper},
\begin{eqnarray}
\label{eq:thermal}
\int 
du 
\frac{ ~ \partial _u g_0(u)
}{ \omega - k v }
\sim 
-\frac{k}{\omega^2}
\left(
1+ 3 \frac{ k^2 \sigma^2}{\omega_0^2}
\right),
\end{eqnarray}
where $\sigma$ is the thermal width of the background electrons' momentum distribution, $g_0(u)$, which we assume here to be non-relativistic, i.e., $\sigma \ll c$.
In deriving Equation \eqref{eq:thermal}, we consider only long wavelengths compared to the local Debye length $\sigma/ \omega_g(x)$, where, the local plasma frequency is $ \omega_g(x)=\sqrt{e^2 n_g(x)/\epsilon_0 m_e}$. 
However, since the longest Debye length is at the minimum of the density, we define it to be $\lambda_D = \sigma/ \omega_0$, and always consider wavelength much longer than the longest local Debye length $\lambda_D = \sigma/ \omega_0$, i.e., $ k^2 \sigma^2/ \omega_0^2 \ll 1$.
Therefore,
\begin{align}
\label{eq:Eevol03}
&
\left(
\frac{ \omega^2  }{\omega_0^2}
-
1
-
3 \frac{ k^2 \sigma^2}{\omega_0^2}
\right)
E_1
+ \epsilon
\left(
1+ 3 \frac{ k^2 \sigma^2}{\omega_0^2}
\right)
\partial^2_k
E_1 
=0
.
\end{align}
If we assume that $\epsilon/k_m^2\leq 1$, where $k_m$ 
is wave number of the most important wave-mode in the system, i.e., $ k_m c/\omega_0 \sim 1$ is the expected fastest unstable mode in the presence of weak pair-beams, 
it ensures that $\epsilon \partial_k^2  E_1 \leq E_1$.
That is, if the inhomogeneity scale is larger than the plasma skin-depth ($c/\omega_0$), we may ignore
$k^2 \sigma^2/\omega_0^2 $ in the second term, and write 
\begin{align}
\label{eq:Eevol05}
&
-
\partial^2_k
E_1
+
\frac{ 3 \sigma^2}{\epsilon \omega_0^2} k^2 
E_1
=
\frac{1}{\epsilon}
\left[
\frac{ \omega^2  }{\omega_0^2}
-
1
\right]
E_1.
\end{align}

\begin{figure}
\includegraphics[width=6.5cm]{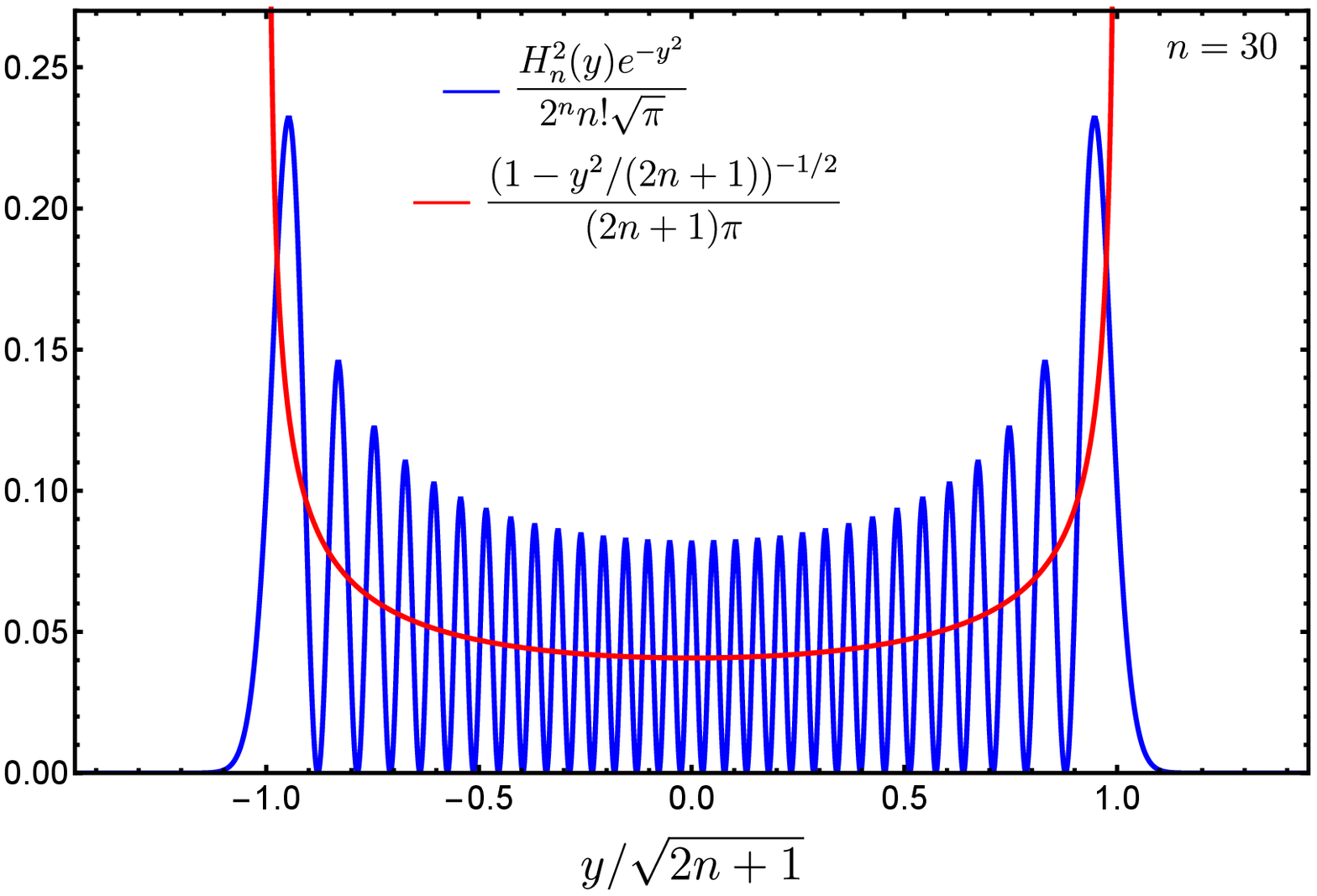}
~
\includegraphics[width=6.5cm]{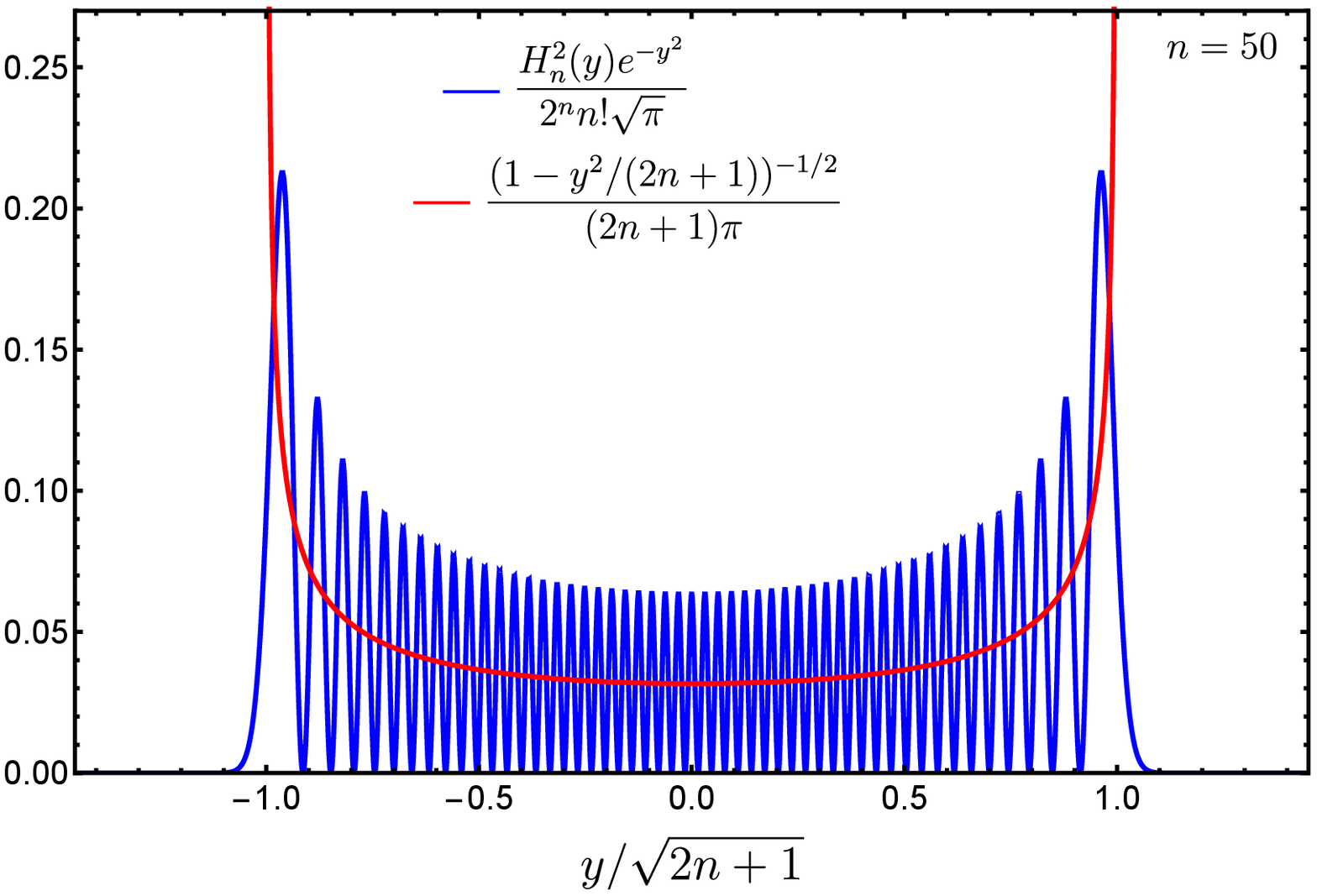}
\caption{
\label{fig:Hn2}
Comparison of the value of $H^2_n\left(y\right)   e^{-y^2 }$ and its approximate form used in Equation~\eqref{eq:dEnapp} for $n=30$ (left) and $n=50$ (right). As $n$ increases the number of oscillations near $y=0$ also increases. 
}
\end{figure}

Equation~\eqref{eq:Eevol05}  has the same structure as the equation for a quantum harmonic oscillator \citep[see, e.g.,][]{shankar_12,griffiths2014}.
If we demand that the solution is finite as $|k| \rightarrow \infty$, we discard the solution of the form $E_1(k,\omega)
\propto e^{+k^2/2 a^2}$.
Thus, the solution is given by
\begin{eqnarray}
\label{eq:Esol}
E_1(k,\omega)
=
A_n  H_n\left(k/a\right) ~ e^{-k^2/2 a^2},
\end{eqnarray}
where, $A_n$ is a normalization constant, $a^4  =   \epsilon \omega_0^2/ (3 \sigma^2)  =  \epsilon k^2_D /(3(2\pi)^2) =  \epsilon /3 \lambda_D^2 $, $k_D$ is the wave number associated with the Debye length, $\lambda_D \equiv \sigma / \omega_0$, ($a$ has the dimension of an inverse length), and
\begin{eqnarray}
\label{eq:disp00}
\frac{1}{\epsilon}
\left[
\frac{ \omega^2  }{\omega_0^2}
-
1
\right]
=
a^{-2} (  2n +1),
~~~~
n \in Z^{+} 
\end{eqnarray}
The basis used in Equation~\eqref{eq:Esol} are written in terms of wavemodes $k$.
However, since the Fourier transform 
\begin{equation}
\mathcal{F}(H_n(k) e^{-k^2/2}) = (-i)^n H_n(x) e^{-x^2/2},
\end{equation}
the structure of the normal modes, in real space and Fourier space, are similar, see Fig.~\ref{fig:Hn2}. Therefore, the solution for each $n$ is given, in real space, by
\begin{eqnarray}
\label{eq:Esolx}
E_1(x,\omega)
=
(-i)^n {\tilde A}_n  H_n\left(x a\right) ~ e^{-a^2 x^2/2 }.
\end{eqnarray}

Note, demanding that the solution remains finite for $|k| \rightarrow \infty$, not only excludes the exponentially divergent part of the solution, but also quantizes the remaining part. That is, for only non-negative integer values of $n$ the solution in Equation~\eqref{eq:Esol} is non-divergent. Thus, the condition in Equation~\eqref{eq:disp00} represents our dispersion relation.
Similar structure of eigenstates was previously found for a quadratic inhomogeneity by~\citet{Ergun+2008}.

Computing the limit of uniform background plasma from the above formulation is more complicated than just taking the limit $\epsilon \rightarrow 0$;
Equation~\eqref{eq:Eevol03}  tells us that in such limit, the dispersion relation is
$  \omega^2/\omega_0^2  - 1 = 3 k^2 \lambda_D^2$.
Since, the dispersion relation in Equation~\eqref{eq:disp00},  can be re-written as
\begin{eqnarray}
\left[
\frac{ \omega^2  }{\omega_0^2}
-
1
\right]
=
\frac{\epsilon}{a^2} (2n +1)
=
\sqrt{ 3 \epsilon \lambda_D^2} (2n +1),
\end{eqnarray}
the limit of uniform background plasma can be obtained via
\[
\epsilon \rightarrow 0,
~~~
n \rightarrow \infty
~~~ 
{\rm such ~ that}
~~~
 \sqrt{ 3 \epsilon \lambda_D^2} (2n +1) \rightarrow 3 k^2 \lambda_D^2 .
\]
By taking the limit in such a way, the solution in Equation~\eqref{eq:Esolx} is reduced to
\[
E_1(x,\omega) \propto \cos \left( k x - n \pi /2\right),
\]
where we used $x a \sqrt{ 2n+1}  \rightarrow x a^2 \sqrt{3 k^2 \lambda_D^2/\epsilon} = k x$.
This is indeed the expected solution in the uniform background case, i.e., the normal modes are Fourier modes rather than Hermite modes.
Here, we have used the large-$n$ limit expansion of the Hermite polynomials, and
we give an explicit expression in such a limit in Equation~\eqref{eq:H2app}.

\section{Adding a weak beam}
\label{sec:WBeam}

We now supplement the inhomogeneous background with a weak plasma beam.
Here, we assume a cold and uniform pair beam, i.e.,
\begin{eqnarray}
f_0^{\pm}
=
n_b \delta(u-u_b),
\end{eqnarray}
where $n_b$ is the uniform number density of the equally dense pair beam, and which is defined in the background plasma frame of reference.

Thus, Equation~\eqref{eq:Eevol01} can be written as
\begin{align}
\label{eq:Eevol06}
&
\left[
k
-
\dfrac{ e^2 n_b}{m_e \epsilon_0 }
\frac{2 k}{ \gamma_b^3 (\omega - k  v_b)^2}
\right]
E_1 (k,\omega)
+
\left[
\omega_{0}^2
\int
du 
\frac{ ~ \partial _u g_0(u)
}{ \omega - k v }
\right]
(1 - \epsilon \partial^2_k )
E_1(k, \omega)
=0
.
\end{align}
Here,  $\gamma_b$ and $v_b$ are the Lorentz factor and the velocity of the pair beam, respectively.
We define $\eta = 2 \alpha/\gamma_b^3$ where $\alpha = n_b/n_0$ is the beam-background density ratio at $x=0$.
Thus, using Equation~\eqref{eq:thermal}
\begin{align}
\label{eq:Eevol07}
&
\left[
1
-
\frac{ \eta  \omega_{0}^2 }{(\omega - k  v_b)^2}
\right]
E_1 (k,\omega)
-
\left[
\frac{ \omega_{0}^2 }{\omega^2 }
\left(
1+ 3 \frac{ k^2 \sigma^2}{\omega_0^2}
\right)
\right]
(1 - \epsilon \partial^2_k )
E_1(k, \omega)
=0
.
\end{align}

This can be rearranged into
\begin{align}
\label{eq:Eevol09}
&
-
\epsilon
\left(
1+ 3 \frac{ k^2 \sigma^2}{\omega_0^2}
\right)
 \partial^2_k 
E_1 
+
\left[
3 \frac{ k^2 \sigma^2}{\omega^2}
+
\frac{ \eta  \omega^2 }{   (\omega - k  v_b)^2}
\right]
E_1 
=
\left(
\frac{ \omega^2 }{\omega^2_{0} }
-
1
\right)
E_1  
.
\end{align}
Again, since $\sigma^2 k^2/\omega_0^2 \ll 1$ and $\epsilon \lambda_D^2 \ll 1 $ the thermal correction term in the first parenthesis can be ignored, and we can recast this equation into an equation for a perturbed quantum harmonic oscillator:
\begin{align}
\label{eq:Eevol10}
&
-
\partial^2_k 
E_1
+
\left[
 \frac{3 \sigma^2}{ \epsilon \omega_0^2} k^2 
+
\frac{  \eta  \omega^2  }{ \epsilon (\omega - k  v_b)^2}
\right]
E_1 
=
\frac{1}{\epsilon}
\left(
\frac{ \omega^2 }{\omega^2_{0} }
-
1
\right)
E_1
.
\end{align}
The small perturbation to the potential by the beam term, i.e., $\eta=2 \alpha/\gamma_b^3 \ll 1$,
means the beam is relativistic and/or dilute.
To compute the change in the dispersion relation, we can use the first order perturbation theory. 
That is, the eigenvalue condition in Equation~\eqref{eq:disp00} becomes
\begin{eqnarray}
\label{eq:disp01}
\frac{1}{\epsilon}
\left[
\frac{ \omega^2  }{\omega_0^2}
-
1
\right]
=
a^{-2} (1+ 2n)
+
\Delta E_n
,
\end{eqnarray}
where,
\begin{eqnarray}
\label{eq:dEn}
\Delta E_n
&=&
\frac{\int dk \dfrac{  \eta  \omega^2   }{ \epsilon (\omega - k  v_b)^2}  H^2_n\left(k/a\right) ~ e^{-k^2/ a^2} }{ \int dk  H^2_n\left(k/a\right) ~ e^{-k^2/ a^2} }
=
\frac{  \eta  \omega^2 /\epsilon }{\int dy ~ H^2_n\left(y\right) ~ e^{-y^2} }
\int dy  \frac{ H^2_n\left(y\right)   e^{-y^2 } }{ ( \omega - y  a v_b)^2   },
~~~~~~
\end{eqnarray}
where $y=k/a$.
It is important to note that, in order to find the modified dispersion relation, we use the first order perturbation theory and explicitly integrate over the Fourier modes labeled by $k$.
Thus the dispersion relation becomes independent of $k$. Instead, it depends on $n$, the label for the eigenmodes (the normal modes) of the system in which the electric field perturbation evolves according to Equation~\eqref{eq:Eevol10}.

Therefore, the full dispersion in presence of a weak beam ($\eta \ll 1$) is given by 
\begin{eqnarray}
\label{eq:fullDisp}
\left[
\frac{ \omega^2  }{\omega_0^2}
-
1
\right]
&=&
\frac{\epsilon}{a^{2}} (1+ 2n)
+
\eta \frac{b^2}{\sqrt{ \pi} 2^n n! }
\int dy  \frac{ H^2_n\left(y\right)   e^{-y^2 } }{ ( y -  b)^2   }  
\end{eqnarray}
where, $b=\omega/a v_b$ and we used 
$ \int dy   H^2_n\left(y\right)   e^{-y^2 } = \sqrt{ \pi} 2^n n! .$

In the following we are interested only in the growth rates, i.e., solution of
Equation~\eqref{eq:fullDisp} with Im$[ \omega]>0$. Therefore, extending the Landau contours of
the integral of Equation~\eqref{eq:fullDisp} to the full complex $\omega$-plane is not needed~\citep{Ferch+1975}.

\subsection{Large-$n$ regime}
\label{sec:large_n}

Here, we approximate the integral in the large-$n$ limit, and also check the regime of  the validity of such an approximation in Appendix~\ref{app:Hermite_approximation}.
We use~\citep{abramowitz+stegun}
\begin{eqnarray}
\label{eq:H2app}
H^2_n\left(y\right)   e^{-y^2 } 
\approx
B_n 
\frac{
\cos^2 \left( y \sqrt{ 2n+1 - \frac{ y^2}{3} } - \dfrac{ n \pi }{2} \right)
}
{
\sqrt{ 
 1-  \frac{y^2}{2n+1}
}
}
~
\Theta \left(  1-  \frac{y^2}{2n+1} \right)
,
\end{eqnarray}
where, $B_n = 2 \left( 2n/e\right)^n$ is a normalization constant and $\Theta (x)$ is the Heaviside step function.
To find a closed form of the dispersion relation in the large-$n$ limit, we need to evaluate the integral in Equation~\eqref{eq:fullDisp}, we average over the oscillatory part of this approximation first, then evaluate the integrals, i.e.,
\begin{eqnarray}
\label{eq:dEnapp}
\epsilon \Delta E_n
&=&
\frac{ 
\eta  b^2 
\int_{-\sqrt{2n+1}}^{\sqrt{2n+1}} dy    \left[ 1- \frac{y^2}{2n+1} \right]^{-\frac{1}{2}}
/ ( b- y   )^2   
}{
\int_{-\sqrt{2n+1}}^{\sqrt{2n+1}} dy \left[ 
 1- \frac{y^2}{2n+1}
\right]^{-\frac{1}{2}} }
=
\frac{\eta b^3 }{[b^2-(2n+1) ]^{3/2}}.
~~~
\end{eqnarray}

Therefore, the dispersion relation is given by
\begin{eqnarray}
\label{eq:disp02}
\frac{ \omega^2  }{\omega_0^2}
-
1
&=&
\frac{ \epsilon (2n+1) }{a^2}
+
\frac{\eta b^3 }{[b^2-(2n+1) ]^{3/2}}
\end{eqnarray}
By defining $ \tilde{\omega} = \omega/\omega_0$, $b_0 = \omega_0/av_b$ such that $b =  \tilde{\omega} ~ b_0  $, it is also convenient to define $r =( 2n+1)/b_0^2$.
Therefore,
\begin{eqnarray}
\label{eq:disp03}
\tilde{\omega}^2-1
&=&
3 \frac{ \sigma^2 }{v_b^2} r  + 
\frac{  \eta  \tilde{\omega}^3  }{ \left(   \tilde{\omega}^2   -  r  \right)^{\frac{3}{2}}},
\end{eqnarray}
where we used 
$a^2 = \epsilon \omega_0^2/3 \sigma^2 a^2$ or, equivalently $\epsilon/a^2 = 3 (\sigma^2/v_b^2) / b_0^2$.

\begin{figure} 
\begin{center}
\includegraphics[width=14.0cm]{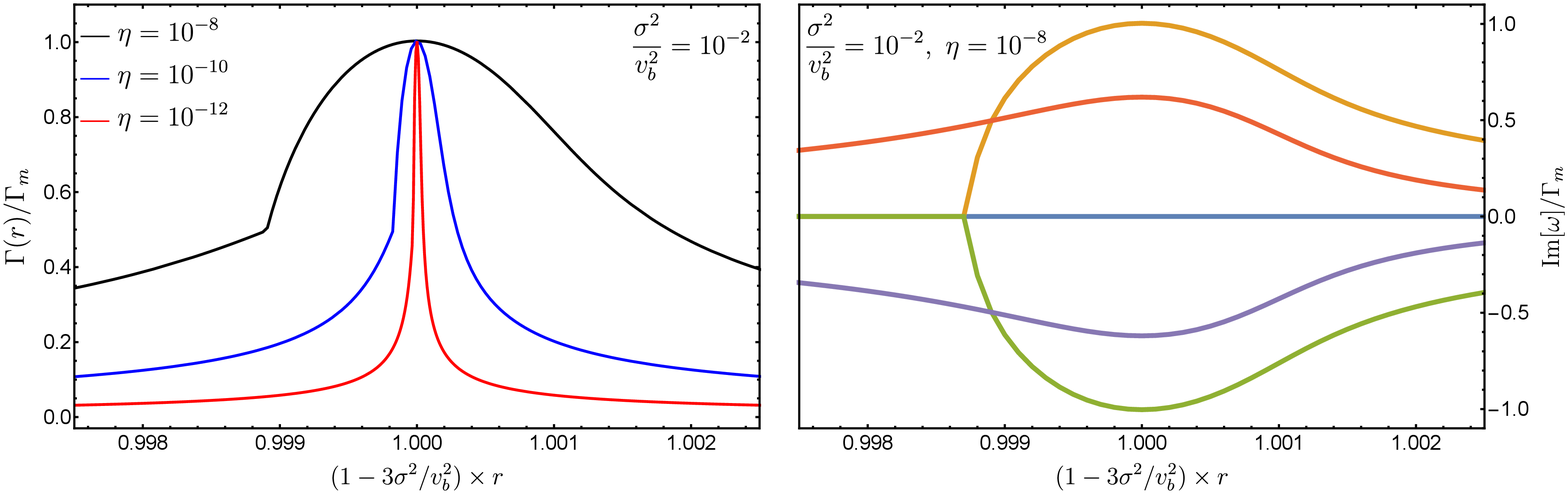}
\end{center}
\caption{
\label{fig:disp}
Numerical solutions of the dispersion relation in the large-$n$ limit.
Left: we show the fastest growth rate obtained by solving Equation~\eqref{eq:disp03} and numerically normalize it to the
fastest growth rate, $\Gamma_m$, given by Equation~\eqref{eq:Gammaf}. 
This is shown for $\sigma^2/v_b^2 = 10^{-2}$, and various values of $\eta$ ($\eta = 10^{-8}, 10^{-10}, \text{ and }10^{-12}$).
Right: we show all solutions of Im$[\omega]$ for the case of $\sigma^2/v_b^2 = 10^{-2}$ and
$\eta = 10^{-8}$.
The right panel shows that the kink features in the fastest growth rate curves of the left panel are  a result of  switching between different unstable branches.
Here $r=(2n+1)/b_0 = (2n+1) v_b a / \omega_0$, and $a^4 = \epsilon \omega_0^2/3 \sigma^2$.
}
\end{figure}

\subsubsection{Fastest growing modes}

When $\eta =0$, i.e., no beam case, the solution of the dispersion relation is 
$
\tilde{\omega}^2
=
\tilde{\omega}^2_0
=
1+ 3 \sigma^2 r / v_b^2$. Since the beam term is such that $\eta \ll 1$, the solution of the full dispersion relation should be such that $\tilde{\omega} = \tilde{\omega}_0 + \delta \tilde{\omega} $, where  $|\delta\tilde{\omega}|\ll\tilde{\omega}_0$. 
Therefore, to lowest order in $\delta\tilde{\omega}$, the dispersion relation can be recast as
\begin{eqnarray}
\tilde{\omega}_0^2 -1 -3 \frac{ \sigma^2 }{v_b^2} r + 
2 \delta \tilde{\omega}
=
2 \delta \tilde{\omega}
&=&
\frac{  \eta  \tilde{\omega}^3  }{ \left(   \tilde{\omega}^2   -  r  \right)^{\frac{3}{2}}}
\approx
\frac{  \eta   }{ \left(   \tilde{\omega}_0^2  + 2 \delta \tilde{\omega} -  r  \right)^{\frac{3}{2}}}
.
\end{eqnarray}
It is easy to show that $ \Im\{\delta \tilde{\omega} \}$ is maximized when $ \tilde{\omega}^2_0 -  r  =0$.
That is, the fastest growing mode occurs at $r=r_m$, and is such that
\begin{eqnarray}
\label{eq:rmax}
r_m 
=1+ 3 \sigma^2 r_m / v_b^2
~~
\Rightarrow
~~
r_m
= 
\frac{1} { 1 - 3 \frac{ \sigma^2 }{v_b^2} }
\approx 
 1 +  3 \frac{ \sigma^2 }{v_b^2}.
\end{eqnarray}
The left panel of Figure~\ref{fig:disp} shows an excellent agreement between $r_m$ and the value of $r$ where the growth rate is maximum when the full dispersion relation is solved numerically.
Therefore, the fastest growth rate is such that
\begin{eqnarray}
2 \delta \tilde{\omega} \sim \frac{\eta }{ (2 \delta \tilde{\omega} )^{3/2} }
\Rightarrow 
\delta \tilde{\omega} \sim \frac{\eta^{2/5}}{2}
\left\{1,
\cos\frac{2\pi}{5}\pm i\sin\frac{2\pi}{5},
\cos\frac{4\pi}{5}\pm i\sin\frac{4\pi}{5}
\right\},
\end{eqnarray}
and the maximum growth rate is
\begin{eqnarray}
\label{eq:Gammaf}
 \Gamma_m 
\sim 
\frac{ \eta^{2/5} }{2}  \sin \frac{2 \pi }{5} ~ \omega_0.
\end{eqnarray}
The computed maximum growth rate in Equation~\eqref{eq:Gammaf} is in an excellent agreement with the fastest growth rate that is found by numerically solving the full dispersion relation near $r=r_m$(see left panel of Figure~\ref{fig:disp}).

\begin{figure}
\begin{center}
~~~~~~
$\xLeftarrow[\textbf{inhomogeneity}]{\textbf{Stronger}}$
~~~~~
$\xRightarrow[\textbf{inhomogeneity}]{\textbf{Weaker}}$
\includegraphics[width=9cm]{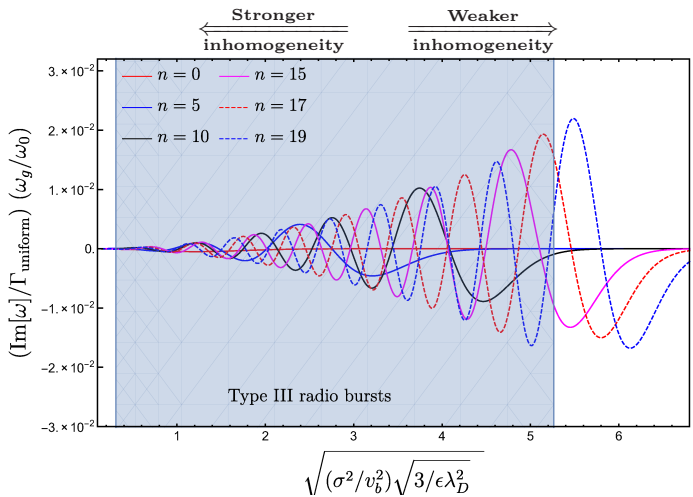}
\end{center}
\caption{
\label{fig:lowndisp012}
Growth rates found by solving the dispersion relation in
Equation~\eqref{eq:fullDisp}, near Im$[\omega] =0$, for various eigenmodes
$n$.
Note, because roots are found near $\omega=\omega_0$, Im[$\omega$] is not necessarily the fastest growth
rate. 
$\Gamma_{\rm uniform}$ is the linear growth rate when the
background plasma is uniform (given by
Equation~\ref{eq:Guniform}).
These solutions are shown for a beam with $3
\sigma^2/v_b^2=10^{-3}$ and $\eta = 10^{-5}$ (parameters relevant for the
inhomogeneities in the type-III radio burst environments).  
The light-blue
shaded region indicates the range of inhomogeneities in these environments, see Section~\ref{sec::solar}.
}
\end{figure}

To compute the eigenmode where the fastest growth occurs $n_m$, we use
\begin{equation}
\label{eq:rmeq}
r_m = \frac{  2 n_m + 1   }{ b_0^2 }= (2n_m + 1 ) \frac{v_b^2 a^2}{\omega_0^2}
=
(2n_m + 1 ) (v_b/\sigma)^2 \sqrt{\frac{\epsilon \lambda_D^2}{3}}
 \approx 1+3 \frac{ \sigma^2 }{v_b^2},
\end{equation}
where, $  \lambda_D = \sigma/\omega_0$.
The fastest growth occurs at
\begin{eqnarray}
2n_m+1
=
b_0^2 
\left( 1   +3 \frac{ \sigma^2 }{v_b^2}  \right)
=
\frac{\sqrt{ 3/ \epsilon \lambda_D^2}}{ (v_b/\sigma )^2 }  
\left(  1+3 \frac{ \sigma^2 }{v_b^2}  \right)
=
\sqrt{ \frac{3}{ \epsilon \lambda_D^2}}
\left(  \frac{ \sigma^2 }{v_b^2}   +3 \frac{ \sigma^4 }{v_b^4}  \right)
.
~~~~~~~~~~~~
\end{eqnarray}
Therefore, the condition to find $n_m$ in the large-$n$ limit, i.e, growth in presence of such an inhomogeneity is (using $\sigma \ll v_b$)
\begin{eqnarray}
\label{eq:condition}
2n_m \gg 0
~~ \Rightarrow ~~
\sqrt{ \frac{ 3}{ \epsilon \lambda_D^2} }
\approx \frac{ \sqrt{ 3}  L_{\rm inh}}{\lambda_D }
\gg \frac{ v^2_b }{ \sigma^2 },
\end{eqnarray}
where $L_{\rm inh}\equiv1/\sqrt{\epsilon}$ is the typical length scale over which the density changes substantially.
It is worth noting that because in the large-$n$ limit, $r_m \sim O(1)$,  the large-$n$ limit is equivalent to the large-$b_0^2$ limit.
The fastest growth rate  of the longitudinal modes, when the background density is uniform, is given by~\citep{Bret-2010-POP,blazari}
\begin{eqnarray}
\label{eq:Guniform}
\Gamma_{\rm uniform} = \dfrac{\sqrt{3}}{2 } \dfrac{  \alpha ^{1/3}}{\gamma_b}
  \omega_g
= 
\dfrac{\sqrt{3}}{2^{4/3}} ~ 
 \eta^{ \frac{1}{3}}   
~ \omega_g,
\end{eqnarray}
where, $\omega_g = \sqrt{ n_g e^2/m_e \epsilon_0}$, is the plasma frequency of the  background electrons in the uniform case that we want to compare to.
Therefore, using Equation~\eqref{eq:Gammaf}, the growth rate in the presence of an inhomogeneity is reduced by a small factor that is  given by
\begin{eqnarray}
\label{eq:reduction}
\frac{\Gamma_m}{\Gamma_{\rm uniform}} = 
\frac{\omega_0}{\omega_g}
\frac{2^{ \frac{1}{3} }   \sin \left( \frac{2 \pi}{5} \right) }{\sqrt{3}}
\eta^{  \frac{1}{15}}.
\end{eqnarray}

\subsubsection{Instability spectral width}

From the numerical solution of  Equation~\eqref{eq:disp03} (see Figure~\ref{fig:disp}), we find that the full-width half max, i.e., the width in $r$ where all the growth is within factor of $0.5$ of the fastest growth rate can be well approximated by
\begin{eqnarray}
\Delta r \sim 2.5 ~ \eta^{2/5}
~~ \Rightarrow ~~
\Delta n = 1.25 ~ \eta^{2/5}  \frac{ \sigma^2 }{v_b^2}   
~
\sqrt{ \frac{3}{ \epsilon \lambda_D^2}}
=
1.25 ~ b_0^2 ~ \eta^{2/5}.
~~~~
\end{eqnarray}
That is, the weaker the beam gets (smaller $\eta$), the slower the fastest growth rate, and the smaller the spectral support around the fastest growing mode, $n_m$.

\begin{center}

\begin{table}
\begin{tabular}{ll}
$n$ ~\hspace{.2cm}~ & ~~  \hspace{1.1cm} ~~ Dispersion relation 
\\
\hline
\hline
0 & ~~
$
\tilde{\omega}^2 -1
=
\dfrac{ 3 \sigma^2}{ v_b^2}  \dfrac{(1)}{b_0^2} 
+ \eta b^2
\left\{ 
\dfrac{\mathcal{I}_0(b)}{\sqrt{\pi}}
\right\} 
$
\\
\hline
1 & ~~
$
\tilde{\omega}^2 -1
=
\dfrac{ 3 \sigma^2}{ v_b^2}  \dfrac{(3)}{b_0^2} 
+ 
\eta b^2
\left\{
2 (b^2-1)  \dfrac{\mathcal{I}_0(b)}{ \sqrt{\pi}} -2
\right\}
$
\\
\hline
2 & ~~
$
\tilde{\omega}^2 -1
=
\dfrac{ 3 \sigma^2}{ v_b^2} \dfrac{(5)}{b_0^2} 
+ 
\eta
 b^2 
\left\{ 
\left(
 4 b^2 \left(b^2-3\right) +5  
\right)
\dfrac{\mathcal{I}_0(b)}{ 2 \sqrt{\pi}}
+
3-2 b^2
\right\}
$
\\
\hline
3 & ~~
$
\tilde{\omega}^2 -1
=
\dfrac{ 3 \sigma^2}{ v_b^2} \dfrac{ (7) }{b_0^2}
+ 
\eta
 b^2 
\left\{ 
\left(2 b^2-3\right) \left(2 b^4-9 b^2+3\right)  \dfrac{\mathcal{I}_0(b)}{3 \sqrt{\pi }}
+
6 b^2-\dfrac{ 4 b^4}{3}-4
\right\}
$
\\
\hline
\end{tabular}
\caption{Low $n$ dispersion relations. Here, $\tilde{\omega}=\omega/\omega_0$, $b=\omega / a v_b = \tilde{\omega}~ b_0$, where $b_0^2=\omega_0^2/a^2 v_b^2 =  (\sigma^2/v_b^2) \sqrt{3/ \epsilon \lambda_D^2 } $, $\mathcal{I}_0(b)$ is given in Equation~\eqref{eq:I0b}, and we used
$\epsilon (2n+1)/a^2= (3 \sigma^2/ v_b^2) [ (2n+1)/b_0^2]  $.
\label{tab:displown}
}
\end{table}

\end{center}

\subsection{Low-$n$ regime}
\label{sec:low_n}

A systematic method to analytically compute the dispersion relations is given in Appendix~\ref{app:Integrals}. Explicit equations for the dispersion relation at $n=0,1,2,3$ are given in Table~\ref{tab:displown}, in terms of 
\begin{equation}
\label{eq:I0b}
\mathcal{I}_0(b) = 
2 \pi  b e^{-b^2} \left[ {\rm Erfi}(b)-i\right] -2 \sqrt{\pi}.
\end{equation}
For parameters relevant for the inhomogeneity in the type-III radio burst environments ($3 \sigma^2/v_b^2=10^{-3}$ and $\eta = 10^{-5}$), we show the roots near $\omega=\omega_0$ of some of these dispersion relations up to $n=19$ in Figure~\ref{fig:lowndisp012}.
The light-blue shaded region in Figure~\ref{fig:lowndisp012} indicates the range of values of inhomogeneities, characterized by $b_0$, in the context of Type-III radio bursts~\citep{Reid+2014}.

The analytical form of the dispersion relation found here are polynomials typically of order $>4$, i.e., for $n>1$, these polynomials multiply $\mathcal{I}_0$ which contain ${\rm Erfi}(b)$. Thus, finding all roots of this dispersion relation is tedious.
To find the fastest growing modes, one would need to solve for all roots of the dispersion relation, and find the solution with the largest growth rate. 
This is a complicated process and we leave this for future work\footnote{Note, the solutions of Figure~\ref{fig:lowndisp012} are roots found near $\omega=\omega_0$, that is Im[$\omega$] is not necessary the fastest growth rates.}.

\subsection{Size of unstable region}
\label{sec:size_n}

An important prediction of the computation of this section is that 
unstable modes are restricted to finite ranges in (position) space. 
That is, if the most unstable state is the eigenmode with $n_m$, the number of peaks,
for modes of the form given by Equation (\ref{eq:Esolx}), is
$n_m+1$. 
The mode and its instability are then restricted to the region between the two outermost peaks (see Fig.~\ref{fig:Hn2} which illustrates the shape of this function).

The width of the unstable region, i.e., the distance between the furthest peaks is $2 x_c$ such that (using Equation~\ref{eq:Esolx})
\begin{eqnarray}
a^2 x_c^2 = 2 n_m +1
~ \Rightarrow ~
x_c^2 = \frac{ (2 n_m +1)  \omega_0^2 }{ v_b^2  a^2} \left( \frac{c }{\omega_0 } \right)^2 \left( \frac{v_b}{c} \right)^2
=
(2 n_m +1)  b_0^2 \left( v_b/c\right)^2  \frac{c^2}{\omega^2_0 }. ~~~~~
\end{eqnarray}
Therefore,
\begin{eqnarray}
\label{eq:xceq}
\frac{ x_c }{c/\omega_0}
=
b_0 ~ \frac{v_b}{c} ~ \sqrt{2 n_m +1 }  
\end{eqnarray}

To facilitate following the application of our computations, in Table~\ref{tab:variables}, we list the most important variables used throughout this work.
The Table also gives various definitions and indications to the significance for some of these variables.

\begin{table}
\begin{tabular}{l l l}
$n_g(x)$  ~ 		& background electron number density profile   ~~  \hspace{1.99cm} ~~ & $n_g(x) = n_0 ( 1 + \epsilon x^2 ) $  \\
$\epsilon$ 			& inhomogeneity parameter  &   \\
$\sigma$ 			& non-relativistic thermal speed of background electron plasma  & $\sigma \ll c$  \\
$a$ 				& --- &  $a^4 = \epsilon_0 \omega_0^2/3 \sigma$\\  
$\omega_g(x)$ 		& local background electron plasma frequency &  $\omega_g(x) = \sqrt{\frac{ e^2 n_g(x)}{m_e \epsilon_0}}$\\  
$\omega_g$ 			& background electron plasma frequency in the uniform case&  $\omega_g = \sqrt{\frac{ e^2 n_g }{m_e \epsilon_0}}$\\  
$\omega_0$ 			& background electron plasma frequency at $x=0$ &  $\omega_0  = \sqrt{\frac{ e^2 n_0}{m_e \epsilon_0}}$\\  
$\lambda_D$ 		& Debye length of background plasma at $x=0$ & $\lambda_D = \sigma/\omega_0$ \\  
$\alpha$ 			& beam-to-background density ratio at $x=0$& $\alpha = n_b/n_0$	\\   
$\eta$ 				& strength of the pair-beam plasma  &  $\eta= 2 \alpha/\gamma_b^2$ 	\\   
$b_0$ 				& strength of inhomogeneity			& $ b_0 = \frac{ \omega_0 }{a v_b} = \sqrt{ \frac{\sigma^2}{v_b^2} \sqrt{\frac{3}{\epsilon \lambda_D^2}}}  $	\\   
$r$ 				&--  & $r=\frac{ 2n+1}{b_0^2}$   	\\   
$r_m$ 				& value of $r$ at which the growth rate is maximum &   $r_m = \frac{1}{(1-3 \sigma^2/v_b^2)}$	\\   
$n_m$ 				& label of the eigenmode with the fastest growth &  --\\   
$x_c$ 				& boundary of the region with expected growth  &   $x_c= b_0 \frac{v_b}{c}  \sqrt{2 n_m +1} \frac{c}{\omega_0}$

\\
\hline
\end{tabular}
\caption{
A list of important variables and definitions used throughout this work.
\label{tab:variables}
}
\end{table}

\section{Comparisons with numerical simulations}
\label{sec:PICsims}

Here, we compare our analytical computations of Section~\ref{sec:WBeam} with PIC simulations of the beam-plasma instability using the SHARP code~\citet{sharp}.

\subsection{Analytical predictions and limitations}
Before presenting our simulations, it is worth noting that all our calculations in this paper assumed that the pair beams are cold. However, 
in order to avoid the known numerical heating \citep[see e.g.,][]{birdsall+1980}, the pair beams are initialized in the simulations with a non-relativistic thermal temperature of $ k_B T_b = 10^{-4} m_e c^2 $ in the beam rest frame.
Thus, we only expect an agreement with our analytical computation for beams moving with relativistic speeds.
For beams that are moving at non-relativistic speeds, additional thermal effects are expected to alter the growth of the unstable modes.

The motivation for our simulations is to compare the results against various predictions of our calculation in Section~\ref{sec:WBeam}.
We list these predictions below:

\begin{enumerate}

\item[1.] Fastest growth rate: it is  practically difficult  to find such a rate in the low-$n$ limit, thus we use the growth rates computed in the large-$n$ limit for reference, i.e., Equation~\eqref{eq:Gammaf}.

\item[2.] A given fastest growth state $n_m$ has $n_m+1$ peaks whose wavelength increases near cut off in real-space, $ \pm x_c$.

\item[3.] For a given fastest growth state $n_m$, the size of growth region, $2 x_c$, is determined by Equation~\eqref{eq:xceq}. This is another prediction from our computation and is independent of whether $n_m$ is computed by solving the dispersion relation or found by counting the number of peaks in the simulation.

\item[4.] For non-relativistic beams,  the thermal effects from the beam-particles are important in the linear regime, and thus, the  evolution is expected to be different (e.g., suppressed) in comparison to our computation that assumes cold beams. 

\end{enumerate}

\subsection{Particle-in-cell simulations}

Here, we present one-dimensional (1D1V) PIC simulations with a quadratic density inhomogeneity for high and low values of 
$b_0 \equiv \sqrt{(\sigma^2/v_b^2) \sqrt{3/\epsilon \lambda_D^2} } \sim 31.675, ~ 3.38, {\rm and ~ }1.49$.
For all simulations, the background plasma is composed of stationary
thermal electron plasma, and a fixed neutralizing background, i.e., simulations are performed in the background plasma frame of reference. The beam-to-background density ratio $\alpha = 0.002$.
Such a low value of $\alpha$ facilitates a direct comparison between 
the results of these simulations to our analytical results in Section~\ref{sec:WBeam}.
For all cases, the initial normalized background number density (for both electrons and the 
fixed-neutralizing background), on a computational domain of length $L$, is given by 
\begin{equation}
\frac{ n(x) }{n_g}
=
\frac
{
1+\epsilon ~(x-L/2)^2
}
{
1+\epsilon ~L^2/12
},
\end{equation}
where $n_g$ is the average number density of the simulated plasmas.
Periodic boundary condition on particles and fields are used, and the pair beams are initially spatially uniform and have a non-relativistic (rest-frame) temperature of $ k_B T_b = 10^{-4} m_e c^2 $.
The level of inhomogeneity in these simulation, which sets the size of the simulation domain, $L$, depends on the velocity of the beam,  $v_b$ and the background electron thermal velocity, $\sigma$.
The inhomogeneity parameter $\epsilon$ in unit of the plasma skin-depth is given by 
\begin{eqnarray}
\epsilon  \frac{ c^2}{\omega_p^2} =  \left( \frac{\omega_0}{\omega_p} \right)^2\frac{ 3 ~(\sigma/c)^2 }{  b_0^4 (v_b/c)^4}.
\end{eqnarray}

In all simulations, we resolve the plasma skin depth by 10 cells, i.e., $\Delta x = 0.1 ~ c/\omega_p$, where $\omega_p$ is the plasma frequency of all simulated species. The time step is fixed and is such that $c \Delta t / \Delta x = 0.4 $. We use a fifth-order interpolation scheme for both, the deposition and back-interpolation steps, which greatly improves the energy conservation of the simulations, see~\citep{sharp} for a more detailed discussion on this issue.

A proper way do study the convergence behavior of PIC simulations,
of such cases, is derived in~\citet{resolution-paper,sharp}.
Such convergence studies, however, go beyond the scope of this
paper.  We here use our simulations {\it only} to demonstrate the
agreement between them and the
calculated linear instability in presence of a quadratic inhomogeneity in the background electron plasma.

\subsubsection{High $b_0$, with relativistic beam: {\rm  Hb0-rel} }
\label{sec:sim_b031}

For this simulation, we initialize electron-positron beam with relativistic speed $v_b/c =  0.99995 $, i.e., $
\gamma_b \sim 100$, the initial background temperature is such that $\sigma^2/v_b^2 = 10^{-2} $.
The pair beams are initialized with a fixed number of 20 particles per cell for each species, while the average number of background electrons per cell is $10^4$.
The level of inhomogeneity is $\epsilon c^2/\omega_p^2  \sim 2.98 \times 10^{-8}$, i.e., a very weak inhomogeneity.
This corresponds to $b_0 \sim 31.675$. 
That is, the growth rate of this simulation is expected to be directly comparable to results found in the large-$n$ limit (see Section~\ref{sec:large_n}).

Therefore, using Equations~\eqref{eq:rmeq}, \eqref{eq:reduction}, and \eqref{eq:xceq}
\begin{equation}
n_m \sim   \frac{b_0^2}{2} (1+3 \frac{ \sigma^2}{v_b^2})    \sim 515,
~~
\frac{ \Gamma_m }{ \omega_0 } \sim  2.08 \times 10^{-4},
~~
\frac{\Gamma_m}{\Gamma_{\rm uniform} } \sim 0.18,
~~ 
2x_c \sim 2047 ~ \frac{c}{\omega_p}.
~~~~~
\end{equation}
Because of this, we choose the box size to be $L = 7500~ c/\omega_p \gg 2 x_c$.

The ratio of the best-fitting growth rate of the potential energy (i.e., $\Gamma_m t \in [3.6,5]$) in our numerical simulation in comparison to the theoretically expected growth rate is $1.22$. This good agreement between the theoretically expected and numerically simulated growth rates is shown in the top panel of Figure~\ref{fig:sims} (red curves).

\subsubsection{Low $b_0$, with relativistic beam: {\rm Lb0-rel} }
\label{sec:sim_b04rel}

In this simulation, we initialize an electron-positron beam that is
moving with relativistic speed $v_b/c = 0.99995 $, i.e., $ \gamma_b \sim 100$,
and the initial background temperature is such that $\sigma^2/v_b^2 = 10^{-3} $.
The pair beams are initialized with a fixed number of 40 particles per cell for each species, while the average number of background electrons per cell is $2 \times 10^4$.
That is, the level of inhomogeneity is $\epsilon c^2/\omega_p^2 \sim 1.16 \times
10^{-5}$, i.e., a strong inhomogeneity.  This corresponds to $b_0=3.38$.

Solutions such as the ones shown in Figure~\ref{fig:lowndisp012} show that the
most unstable eigenmode is $n_m=9$, thus the expected number of peaks during
the linear evolution in the charge density is $10$.
the region where such growth is given by Equation~\eqref{eq:xceq}; $x_c = 20.69
c/\omega_p$.

Excellent agreement between the predicted number of peaks and the
size of the growth region is show in the bottom panel of Figure~\ref{fig:sims}.
Moreover,  
the ratio of the best-fitting growth rate of the potential energy (i.e., $\Gamma_m t \in [3.6,5]$) in our numerical simulation in comparison to the theoretically expected growth rate is $1.2$. That is, we see a good agreement between the theoretically expected (large-$n$ limit) and numerically simulated growth rates of the simulation.
This is shown in the top-left
panel of Figure~\ref{fig:sims} (blue curves).

\subsubsection{Low $b_0$, with non-relativistic beam: {\rm  Lb0-nonrel} }
\label{sec:sim_b04nonrel}

In this simulation, we initialize an electron-positron beam  that is
moving at non-relativistic speed $v_b/c = 0.1 $, and the initial background
temperature is such that $\sigma^2/v_b^2 = 0.099$.
The pair beams are initialized with a fixed number of particles per cell of $10^3$ per species, while the average number of background electrons per cell is $5 \times 10^5$.
That is, the level of
inhomogeneity is $\epsilon c^2/\omega_p^2 \sim 0.084$, i.e., a very strong
inhomogeneity.  This corresponds to $b_0=1.49$.

\begin{figure} 
\includegraphics[width=13.2cm]{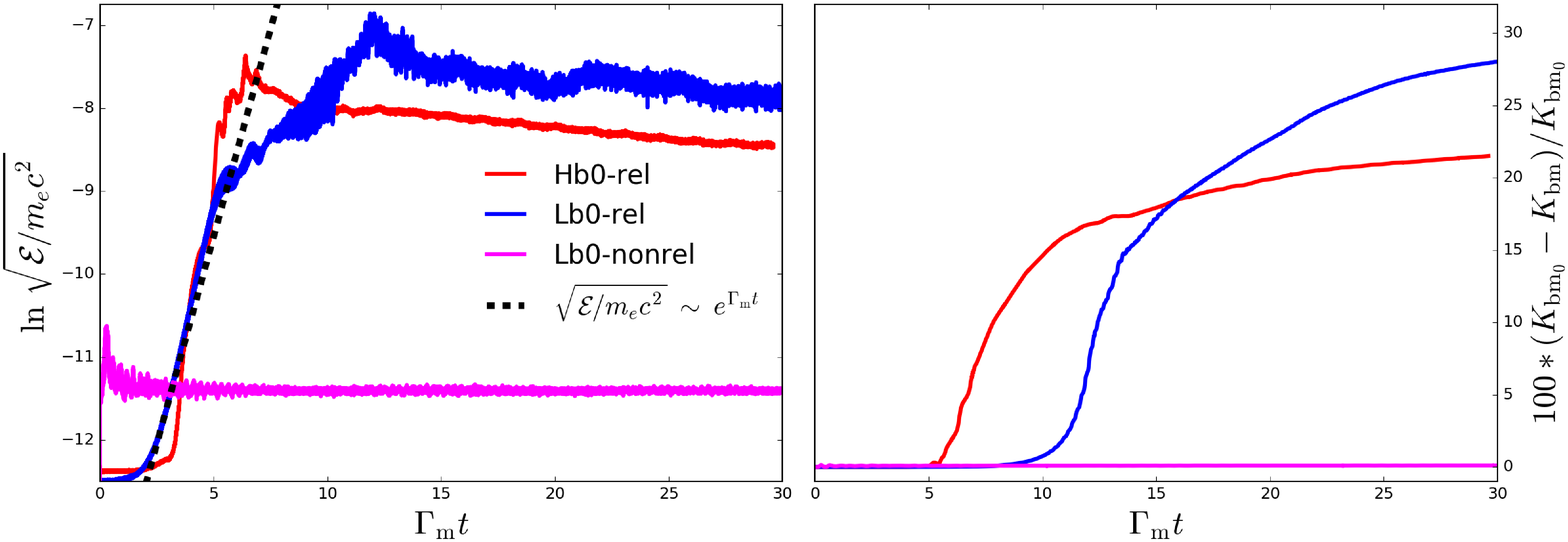}
\includegraphics[width=12.6cm]{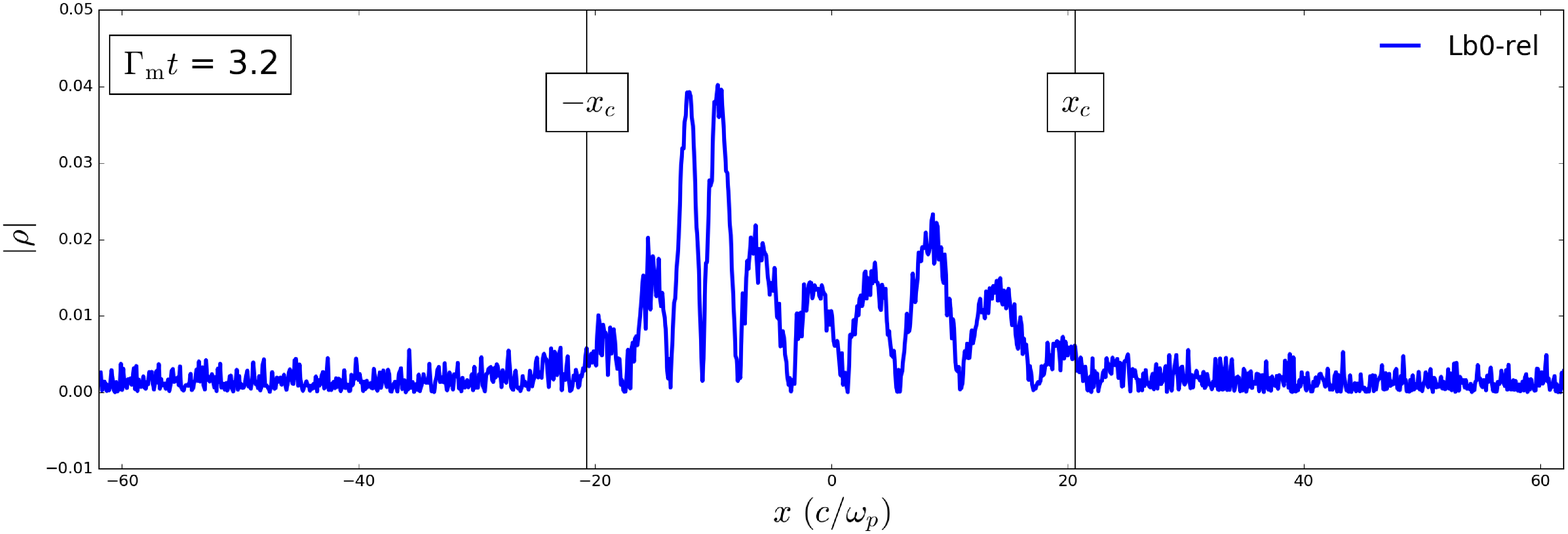}
\caption{
\label{fig:sims}
Particle-in-cell simulation results.
Top Left: Growth of the potential energy density per computation particle, $\mathcal{E}$, (normalized to $m_e c^2$), in various simulation.
The time is normalized to the expected growth rate in the large-$n$
limit $\Gamma_{\rm m }$, i.e., given in Equation~\eqref{eq:Gammaf}.
For {\rm Lb0-nonrel}, the time is further divided by a factor of $100$.
Top Right: The evolution of percentage energy loss by beam particles in various simulations.
Bottom: the absolute value of the charge density on the grid at $\Gamma_m t \sim 3.2$, i.e., near the end of the linear regime potential energy growth (top left figure) of the {\rm Lb0-rel} simulation. 
Since the unstable modes are traveling along the beam direction ($+x$-direction), their reflection~\cite[see, e.g., Figure 4 of][]{sim_inho_18} at higher density regions, i.e., $|x|>0$, results in asymmetric structure 
shown in the bottom panel.
}
\end{figure}

Naive application of our results above suggest a nontrivial growth rate, which is not seen in the numerical calculation.  
We attribute this to the violation of the cold beam approximation in our analytic calculation and suggest that thermal effects of the beam-particle momentum
distribution almost completely suppress the growth in such a case (magenta curves in
Figure~\ref{fig:sims}).

\section{Applications }
\label{sec:application}

Here, we apply the results of Section~\ref{sec:WBeam}, to
astrophysical plasmas within various astrophysical contexts that span many
scales while adhering to its limitations found in Section~\ref{sec:PICsims}. 
This is done with the goal of determining whether the inhomogeneity, 
with the structure studied here, can suppress the growth of the unstable wave-modes.

\subsection{Beam-plasma instabilities in the intergalactic medium (IGM)}

TeV-photons emitted by blazars
create via pair production very energetic pair beams that propagate through 
the ionized intergalactic medium (IGM)~\citep{blazari,blazarii,blazariii,paperiv,paperv}.
Fermi-LAT observations at GeV energies show that
the expected GeV photons
that result from the inverse Compton cascade of these pair beams 
on cosmic microwave photons, are
missing~\citep{bowtiesi,bowtiesii,bowtiesiii,Fermi_aligned+2018,broderick+2018}.
A plausible explanation of such a mystery is that virulent kinetic plasma
instabilities in the IGM, induced by the pair
beams, reduce the pair-beam energy on time scales much shorter than that of the
inverse Compton cascade.
The validity of such a scenario strongly depends on
the non-linear saturation of these
instabilities~\citep{Miniati-Elyiv-2013,sironi+giannios+2014,chang:2014,Kempf+2016,linear-paper,resolution-paper,Vafin+2018,Vafin+2019}.

It was suggested by~\citet{Miniati-Elyiv-2013} that the inhomogeneity in the IGM
number density can potentially suppress the growth of such instabilities.
However, it was demonstrated with PIC simulations that the condition for
suppressing the instabilities computed in \citet{Miniati-Elyiv-2013} is invalid,
and cannot suppress even the slowest type of instabilities in such systems,
i.e., the longitudinal instability~\citep{sim_inho_18}.

Our assumption of a fixed background is exceedingly well justified in voids within the IGM.  The dynamical time over which gravitational instabilities will modify inhomogeneities in low-density regions is greater than $10^{11}$~years.  In comparison, estimates for the typical growth times for blazar-driven beam-plasma instabilities range from $10^3$-$10^5$~years~\citep{blazari}.  As we will see below, these esimates are not substantially changed, and thus over many growth times a fixed background is an excellent approximation.

Below, we use our computed growth rates to demonstrate that the level of inhomogeneity in the IGM (for
inhomogeneities of the structure studied in this work) is indeed not sufficient to
suppress the longitudinal instability driven by the pair beams in the IGM.
The relevant parameters for such a situation are $v_b \sim c$, and the background
temperature of electrons of the IGM is such that $\sigma/c \sim \sigma/v_b = 7.5
\times 10^{-3}$.  The inhomogeneity scale length is $L_{\rm inh} \approx
10^2-10^3$ kpc 
at mean density~\citep{Miniati-Elyiv-2013}.  The Debye length is
$\lambda_D \sim 84$ km. Thus
\begin{eqnarray}
\frac{\sqrt{ 3} ~ L_{\rm inh} }{\lambda_D}
\approx 6.4\times (10^{16} -10^{17})
~,~~~ ~~
\frac{\sigma^2}{v^2_b} \approx 5.6 \times 10^{-5}
\end{eqnarray}
Thus the conditions underlying the analysis in Section~\ref{sec:WBeam} are satisfied.
The index of the fastest growing wavemode, using Equation~\ref{eq:condition}, is given
\begin{eqnarray}
n_m
&=&
\sqrt{\frac{3 }{4} \frac{L_{\rm inh}}{\lambda_D}}
\left(  \frac{ \sigma^2 }{v_b^2}   +3 \frac{ \sigma^4 }{v_b^4}  \right)
-\frac{1}{2}
\approx
(1-3)\times 10^{4} ,
~~~~~~~
\end{eqnarray}
placing the blazar-driven beam plasma instabilities well within the large-$n$ regime.  
For the longitudinal modes we studied here, $\eta = \alpha/\gamma_b^2$, however for the blazar-driven beam-plasma instabilities, the oblique modes are the fastest unstable linear modes for which $\eta = \alpha/\gamma_b$
\citet{bret2010}. 
The typical parameters for these instabilities are $\alpha = 10^{-16}$ and $\gamma_b = 10^6$, thus, the expected reduction to the growth rate is roughly $10^{\frac{-16-6}{15}} \sim 1/40$.
As a result, the inhomogeneity is unlikely to suppress the linear growth of the blazar-driven beam-plasma instability in the IGM in the cold-beam limit.

\subsubsection*{Application to  the ``cosine'' simulation of~\citet{sim_inho_18}  }

Here, we show how our analytical results compare to a PIC simulation with an inhomogeneity that is
comparable to the one considered in this work: in ~\citet{sim_inho_18}, PIC
simulations using the SHARP code~\citep{sharp} have shown that the growth of the
instability persists (albeit at slightly lower rates) in the presence of a very
strong inhomogeneity.

The ``cosine'' simulation of \citet{sim_inho_18} has a background inhomogeneity
that varies as a cosine with minimum at the center of the simulation box (see
Figure 1 of \citealt{sim_inho_18}).
Near the minimum of the cosine,
$n(x)/n_g   \approx 0.9 + \pi^2 (x/L)^2/5$, that is $n_0 = 0.9 n_g$, and $\epsilon = \pi^2/(5 n_0 L^2 )$.
We can test the computation presented above against the results of this simulation.
The ``cosine'' simulation had the following numerical parameters:
\begin{eqnarray}
&&
2 \alpha=0.002/0.9 = 0.0022
~,~
\gamma_b=100
~,~
\omega_0 = \sqrt{0.9 }~ \omega_g
~
\Rightarrow
~
\eta = \frac{2 \alpha}{\gamma_b^3} = 2.22 \times 10^{-9},
\nonumber 
\\
&&
L = 128 c/\omega_p
~,~
\sigma^2=3\times10^{-4} c^2
~
\Rightarrow
~
b_0 \sim 1.697.
~~~~~~~
\end{eqnarray}
Therefore, the predicted and simulated reduction (see Table 1 of \citealt{sim_inho_18}) in the linear growth rate due to the inhomogeneity is given by
\begin{eqnarray}
\left( \frac{  \Gamma_m }{\Gamma_{\rm uniform}}  \right)_{\rm predicted}
&=&
\frac{\omega_0}{\omega_g}
\frac{2^{ \frac{1}{3} }~ \sin \left( \frac{2 \pi}{5} \right) }{\sqrt{3}}
\eta^{  1/15}
=
0.174,
\nonumber \\
\left( \frac{  \Gamma_m }{\Gamma_{\rm uniform}}  \right)_{\rm simulation}
&=&
0.2.
\end{eqnarray}

That is, our computed reduction in the growth rate is in very good agreement with the growth rate of the ``cosine'' simulation of \citet{sim_inho_18}.
Moreover, another prediction of the computation presented in this work is an
importance characteristic of the growing modes (Hermite basis with $n \sim
n_m$).
That is, the characteristic wavelength of the fastest growing mode,
just before the region where it is no longer supported (near $y^2=2n+1$),
is larger in comparison to the
wavelength near $y=0$.  This is consistent with the structure shown close to the
end of the linear growth phase of the ``cosine'' simulation shown in Figure 3
(third panel) of \citet{sim_inho_18}.

\subsection{Type-III solar radio bursts}
\label{sec::solar}

Type-III solar radio bursts are the most prolific type of solar radio
burst~\citep{Reid+2014}.  It is generally accepted that during these bursts,
solar electrons are accelerate  following a
reconfiguration of coronal magnetic field lines,
which converts magnetic field energy into kinetic energy.
A theory to describe
type-III burst was first developed by~\citet{Ginzburg+1958}.
They assume that a longitudinal beam-plasma instability, which is driven by
the electron beams, generates Langmuir waves
at the local plasma frequency, and the electromagnetic emission is 
a  result of various scatterings and wave decay processes of these
Langmuir waves. 
The scattering of Langmuir waves results in emission at the
fundamental plasma frequency, while wave decay results in emission at the second
harmonic, i.e., twice the local plasma frequency~\citep{Melrose+2009}.
In situ measurements at 1 AU, show a clear sign of plasma wave energy above
background thermal noise, and the observed particle-momentum distributions of
the electrons do not show the plateau distribution predicted from quasi-linear
theory for instabilities operating in homogeneous or weakly
inhomogeneous background plasmas
\citep{Vedenov-1967,Lin+1981}.

For type-III  radio bursts, the electron beams and plasmas observed at 1 AU, have the following characteristics (see,  e.g., Ref. \citet{Ergun+98,krafft-2013}).
\begin{eqnarray}
&&
\frac{ L_{\rm inh} }{\lambda_D}
\sim
300-2000
~,~~~ ~
\frac{v_b}{c}  \sim 0.05- 0.3
~,~~~ ~
k_B T_e \sim 10 { ~ \rm eV}
\nonumber \\
&&
\Rightarrow
~~~
\frac{\sqrt{3} ~ L_{\rm inh}}{\lambda_D} 
\sim 520-3465
~,~~~ ~~
\frac{v^2_b}{\sigma^2}
\sim
128-4600.
\end{eqnarray}
This implies a wide range of values for $b_0$
\begin{eqnarray}
b_0 = \frac{\omega_0}{av_b} =  \sqrt{ \frac{\sigma^2}{v_b^2} \sqrt{\frac{3}{\epsilon \lambda^2_D}}}
\sim
0.3-5.3.
\end{eqnarray}
For a typical electron density of $n_0\sim10~{\rm cm^{-3}}$ within the solar wind and beam density ratios $\alpha \sim 10^{-5}$, the implied instability growth rate is of order $10^{-4}$-$10^{-3}$ seconds~\citep{krafft-2013}.
In comparison, the typical timescale over which the inhomogeneous structures evolve in the solar wind is $ \sim  {\rm AU} / (500 ~{\rm km/sec} )\sim 3 $ days.  Thus, again, our ansatz of a fixed background is exceedingly well justified.

The light-blue shaded region in Figure~\ref{fig:lowndisp012}, shows this range.
While, for a quadratic inhomogeneity, our calculation here shows that the
inhomogeneity slows the growth of the beam-plasma longitudinal mode, leading to
a suppression of the growth rate by a factor of $\sim 7$ for $b_0=3.38$ 
(Section~\ref{sec:sim_b04rel}). 
However, our PIC simulation of Section~\ref{sec:sim_b04nonrel} shows that
there is almost a complete suppression of the instability, and the beam looses only 0.1\% of its initial energy, as show in Figure~\ref{fig:sims}, when $b_0 \sim 1.5$.
As discussed above, this is most likely due to thermal effects in the beam plasma.

\section{Discussion and Conclusions}
\label{sec:conclusion}

In this paper, we study the linear evolution of beam-plasma systems, in a
fully-relativistic setting, starting form the linearization of the kinetic
equations in one dimension, i.e., linearization of the Vlasov-Poisson equation.
Unlike previous studies, we do not follow the evolution of pre-existing Langmuir waves, instead we 
focus on how the waves are
excited due to the propagation of the beam, and calculate their growth rates.

We derive a novel analytical formula for the growth rate of the
longitudinal instability; see
Equation~\eqref{eq:Gammaf}. This is formally valid only in the large-$n$ limit
(week inhomogeneity limit). However, as shown in Section~\ref{sec:PICsims}, this
formula also provides a good agreement with the growth
rate in a simulation with strong inhomogeneity, i.e., where the most unstable
eigenstate is $n_m=9$ (Section~\ref{sec:sim_b04rel}).  Another important
implication of our computation is that, in the cold-beam limit, the reduction in
the growth rate is independent of the level of inhomogeneity and only depends on
the beam strength $\eta=\alpha/\gamma_b^3$.
As we discuss in Section~\ref{sec:NoBeam}, the limit of uniform background plasma, can not be obtained by simply taking $\epsilon \rightarrow 0$.
That is, the correct normal modes  are Fourier modes instead of the Hermite modes, in which case, trivially, there is no reduction in the growth rate.

The strength of the
inhomogeneity, i.e., the value of $b^2_0 = (\sigma^2/v_b^2) \sqrt{3/\epsilon
  \lambda_D^2}$, determines the most unstable eigenstate, i.e., the structure of
the unstable modes and the size of the linearly unstable region.
Including the effect of finite beam temperatures is important for studying the
stability of systems with beam particles moving  at non-relativistic speeds, e.g.,
propagating beams of type-III radio bursts.  This leads to suppression of
expected growth, i.e., in the cold-beam limit, as seen in
Section~\ref{sec:PICsims}.  This can be done analytically using the same
procedure followed here. However, computing the resulting dispersion relation in
this case is much more complicated and we leave this to future work.

\acknowledgments

\section*{Acknowledgments}

We would like to thank Paul Tiede for participating in various discussions related to this manuscript.
M.S., C.P., and E.P. acknowledge support by the European Research Council under ERC-CoG grant CRAGSMAN-646955.
A.E.B.~is supported in part by a grant from the Delaney family, by Perimeter Institute, and by the Natural Sciences and Engineering Research Council of Canada through a Discovery Grant. Research at Perimeter Institute is supported in part by the Government of Canada through the Department of Innovation, Science and Economic Development Canada and by the Province of Ontario through the Ministry of Colleges and Universities.
P.C. is supported by the NASA ATP program through NASA grant NNH17ZDA001N-ATP. 
A.L. receives financial support from the Programme National des Hautes Energies (France)

\appendix

\section{Computing the dispersion relation for low $n$ }
\label{app:Integrals}

As assumed throughout the paper, 
we define  $\tilde{\omega}=\omega/\omega_0$,  $b = \omega/a v_b =\tilde{\omega} b_0$,
\begin{eqnarray}
\epsilon /a^2 
=
 3 ( \sigma^2/v_b^2)/ b^2_0
,
~{\rm and}~
b_0^2 
= 
\frac{\omega_0^2}{a^2v_b^2}
=
 \frac{\sigma^2}{v_b^2}  \sqrt{ \dfrac{3}{\epsilon \lambda_D^2 } }.  
~~~~~~~~
\end{eqnarray}

In order to find the explicit form of the dispersion relation, i.e.,
Equation~\eqref{eq:fullDisp}, we need to compute integrals of the form
\begin{eqnarray}
\mathcal{I}_l 
=
\int^{\infty }_{- \infty}
 \frac{(y^2)^l e^{-y^2} dy }{(y-b)^2}.
\end{eqnarray}
We are only interested in growth rates, i.e., solutions of
Equation~\eqref{eq:fullDisp} with Im$[b]>0$. Therefore, extending the Landau
contours of the integral of Equation~\eqref{eq:fullDisp} to the full complex
$\omega$-plane is not needed~\citep{Ferch+1975}.  For $l=0$, the integral is
given by
\begin{eqnarray}
\label{eq:I0}
\mathcal{I}_0
&=&
\int^{\infty }_{- \infty}
 \frac{e^{-y^2} dy }{(y-b)^2}
=
2 \pi  b e^{-b^2} \left[ {\rm Erfi}(b)-i\right] -2 \sqrt{\pi}
\nonumber \\
&&
~~~ \hspace{.65cm} ~~~~
=
- 2 
\left[
i \pi  b e^{-b^2} \left[ {\rm Erf}(ib)+1\right] + \sqrt{\pi}
\right]
~~~~~~~~
\end{eqnarray}
where the complex error function, Erfi$(b) \equiv -i $ Erf($ib)$, is defined in terms of the error function Erf.
The integral, $\mathcal{I}_0$, is related to the commonly used plasma dispersion function
\begin{equation} 
Z(b)
 \equiv 
\int^{\infty }_{- \infty}
  \frac{dy}{ \sqrt{ \pi }} ~ \frac{ e^{-y^2} }{y-b}
~~ \Rightarrow ~~
\mathcal{I}_0 (b) =  \sqrt{\pi} Z^{'}(b).
\end{equation}
To compute $\mathcal{I}_l$ ($l=1,2, \dots $), we define
\begin{eqnarray}
\mathcal{I}_0^{h}
&=&
\int^{\infty }_{- \infty}
 \frac{e^{-(1+h)y^2} dy }{(y-b)^2}
 =
(1+h)^{\frac{1}{2}} ~ \mathcal{I}_0( \sqrt{1+h } ~b ),
~~~~~~~~
\end{eqnarray}
where, $-1<h <1$. Therefore,
\begin{eqnarray}
\label{eq:Ilforms}
\mathcal{I}_l = (-1)^l \lim_{h \rightarrow 0^{+}} \frac{d^l   \mathcal{I}^{h}_0}{dh^l}
=
b^{2(l-1)} \left(b^2-l \right) \mathcal{I}_0(b)+ \frac{  \sqrt{\pi} }{2^{l-1}} f_l(b)
~~~~~~~
\end{eqnarray}
where $f_l(b)$ are polynomials of $b$ whose explicit forms can be trivially derived using Equation~\eqref{eq:Ilforms}.
The explicit forms of $f_l(b)$, for $l=1,2 \dots 9$, are given in Table~\ref{tab:flforms}.
With the help of the above integrals, and the explicit form for the Hermite polynomials $H_n(y)^2$, 
an explicit computation for the dispersion relation for all $n$ is possible. 
However, it becomes progressively complicated at large-$n$ to find its roots.
Below we present the computation of the dispersion relations for $n=0,1,2,3$.

\subsection{$n=0$}
Here, the integral we need to compute is
\begin{eqnarray}
\frac{\int dy  \frac{   e^{-y^2 } }{ ( y -  b)^2   } }{\int dy ~ H^2_0\left(y\right) ~ e^{-y^2} }
=
\frac{\mathcal{I}_0(b)}{\sqrt{\pi}}.
~~~
\Rightarrow
~~~
\left[
\frac{ \omega^2  }{\omega_0^2}
-
1
\right]
=
\frac{\epsilon}{a^{2}}
+ \eta b^2
\frac{\mathcal{I}_0(b)}{\sqrt{\pi}}.
\end{eqnarray}

\subsection{$n=1$}

The integral we need to compute is
\begin{eqnarray}
\frac{\int dy  \frac{ 4 y^2   e^{-y^2 } }{ ( y -  b)^2   } }{\int dy ~ H^2_1\left(y\right) ~ e^{-y^2} }
=
\frac{2 \mathcal{I}_1}{ \sqrt{\pi}}
=
2 (b^2-1)  \frac{\mathcal{I}_0}{ \sqrt{\pi}} -2.
\end{eqnarray}
Therefore, the dispersion relation for $n=1$ is given by
\begin{eqnarray}
\left[
\frac{ \omega^2  }{\omega_0^2}
-
1
\right]
=
\frac{3\epsilon}{a^{2}}
+ \eta b^2
\left[ 
2 (b^2-1)  \frac{\mathcal{I}_0}{ \sqrt{\pi}} -2
\right].
\end{eqnarray}

\subsection{$n=2$}

The integral we need to compute is
\begin{eqnarray}
&&
\frac{\int dy  \frac{ \left(4 y^2-2\right)^2   e^{-y^2 } }{ ( y -  b)^2   } }{\int dy ~ H^2_2\left(y\right) ~ e^{-y^2} }
=
\frac{\mathcal{I}_0- 4 \mathcal{I}_1 + 4 \mathcal{I}_2}{ 2 \sqrt{\pi}}
=
3-2 b^2 +
\left[ 4 b^2 \left(b^2-3\right) + 5\right]    \frac{\mathcal{I}_0 }{2 \sqrt{\pi}}.
~~~~
~~~
\end{eqnarray}
Therefore, the dispersion relation for $n=2$ is given by
\begin{eqnarray}
\left[
\frac{ \omega^2  }{\omega_0^2}
-
1
\right]
=
\frac{5\epsilon}{a^{2}}
&+& \eta
 b^2 \left(3-2 b^2\right) 
+
\eta b^2
\left[
2 b^2 \left(b^2-3\right) + \frac{5}{2}
\right]
\frac{\mathcal{I}_0}{ \sqrt{\pi}}.
~~~~~~~
\end{eqnarray}

\subsection{$n=3$}

The integral we need to compute is
\begin{eqnarray}
\frac{\int dy  \frac{  \left(8 y^3-12 y\right)^2   e^{-y^2 } }{ ( y -  b)^2   } }{\int dy ~ H^2_3\left(y\right) ~ e^{-y^2} }
&&
\frac{8 \mathcal{I}_3 -24 \mathcal{I}_2 + 18 \mathcal{I}_1   }{ 6 \sqrt{\pi}}
=
-\frac{4 b^4}{3}+6 b^2-4
+
\left[ 4 b^6-24 b^4+33 b^2-9 \right]    \frac{\mathcal{I}_0 }{3\sqrt{\pi}}.
\nonumber \\
\end{eqnarray}
Therefore, the dispersion relation for $n=3$ is given by
\begin{eqnarray}
\left[
\frac{ \omega^2  }{\omega_0^2}
-
1
\right]
=
\frac{7\epsilon}{a^{2}}
&+& \eta
 b^2 \left(-\frac{4 b^4}{3}+6 b^2-4\right) 
+
\eta b^2
\left[
4 b^6-24 b^4+33 b^2-9
\right]
\frac{\mathcal{I}_0}{ 3 \sqrt{\pi}}.
~~~~~~~
\end{eqnarray}

\begin{table}
\caption{
Explicit form of $f_l(b)$, for $l=1,2, \dots, 9$, used to define $\mathcal{I}_l$ in Equation~\eqref{eq:Ilforms}.
\label{tab:flforms}
}
\begin{tabular}{l|l}
\hline
$l$  \hspace{0.3cm} & ~\hspace{1.0cm} ~ $f_l(b) $
\\ \hline
$ 1 $ & $-1 $ \\
$ 2 $ & $1-2 b^2 $ \\
$ 3 $ & $-4 b^4+6 b^2+3 $ \\
$ 4 $ & $-8 b^6+20 b^4+18 b^2+15 $ \\
$ 5 $ & $-16 b^8+56 b^6+60 b^4+90 b^2+105 $ \\
$ 6 $ & $2 \left(-16 b^8+72 b^6+84 b^4+150 b^2+315\right) b^2+945 $ \\
$ 7 $ & $-64 b^{12}+352 b^{10}+432 b^8+840 b^6+2100 b^4+5670 b^2+10395 $\\
$ 8 $ & $2 \left(2 \left(2 \left(2 \left(-8 b^6+52 b^4+66 b^2+135\right) b^2+735\right) b^2+4725\right) b^2+31185\right) b^2+135135 $ \\
$ 9 $ & $2 \left(2 \left(2 \left(2 \left(2 \left(-8 b^6+60 b^4+78 b^2+165\right) b^2+945\right) b^2+6615\right) b^2+51975\right) b^2+405405\right) b^2+2027025 $ \\
\end{tabular}
\end{table}

\section{Approximating the Integral in Equation~[\ref{eq:fullDisp}]}
\label{app:Hermite_approximation}

To quantitatively evaluate the accuracy of our approximation in Equation~\eqref{eq:H2app}, we can first compute how fast it can recover the normalization as $n$ increases,
\begin{equation}
N = \int H_n^2(y) dy e^{-y^2} = 2^n   n!  \sqrt{\pi}.
\end{equation}
To compute the approximate value of this normalization $N$, as done in Section~\ref{sec:large_n}, we average over the oscillatory part of this approximation first, namely
\begin{eqnarray}
\tilde{N} 
&=& 
2 \left( 2n/e\right)^n 
\int_{-\sqrt{2n+1}}^{ \sqrt{2n+1} }  dy
\frac{
\cos^2 \left( y \sqrt{ 2n+1 - \frac{ y^2}{3} } - \dfrac{ n \pi }{2} \right)
}
{
\sqrt{ 
 1-  \frac{y^2}{2n+1}
}
}
\approx
 \left( 2n/e\right)^n 
\int_{-\sqrt{2n+1}}^{ \sqrt{2n+1} } 
\frac{
dy}
{
\sqrt{ 
 1-  \frac{y^2}{2n+1}
}
}
\nonumber \\
&=&
\pi  \sqrt{2 n+1} \left(\frac{2 n}{\exp (1)}\right)^{n}.
\end{eqnarray}

\begin{figure}
\centering
\includegraphics[height=4.cm]{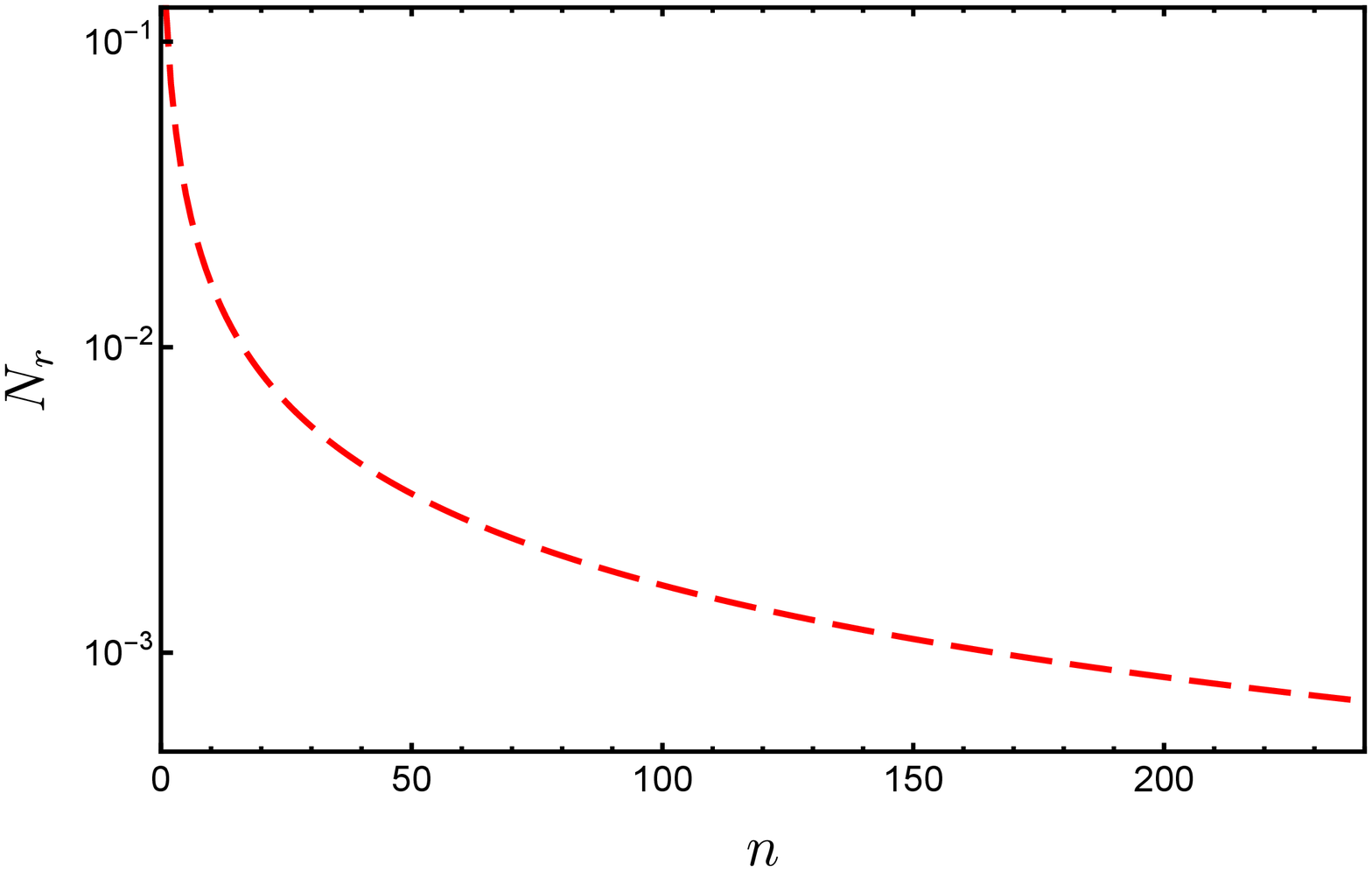}
~
\includegraphics[height=4.1cm]{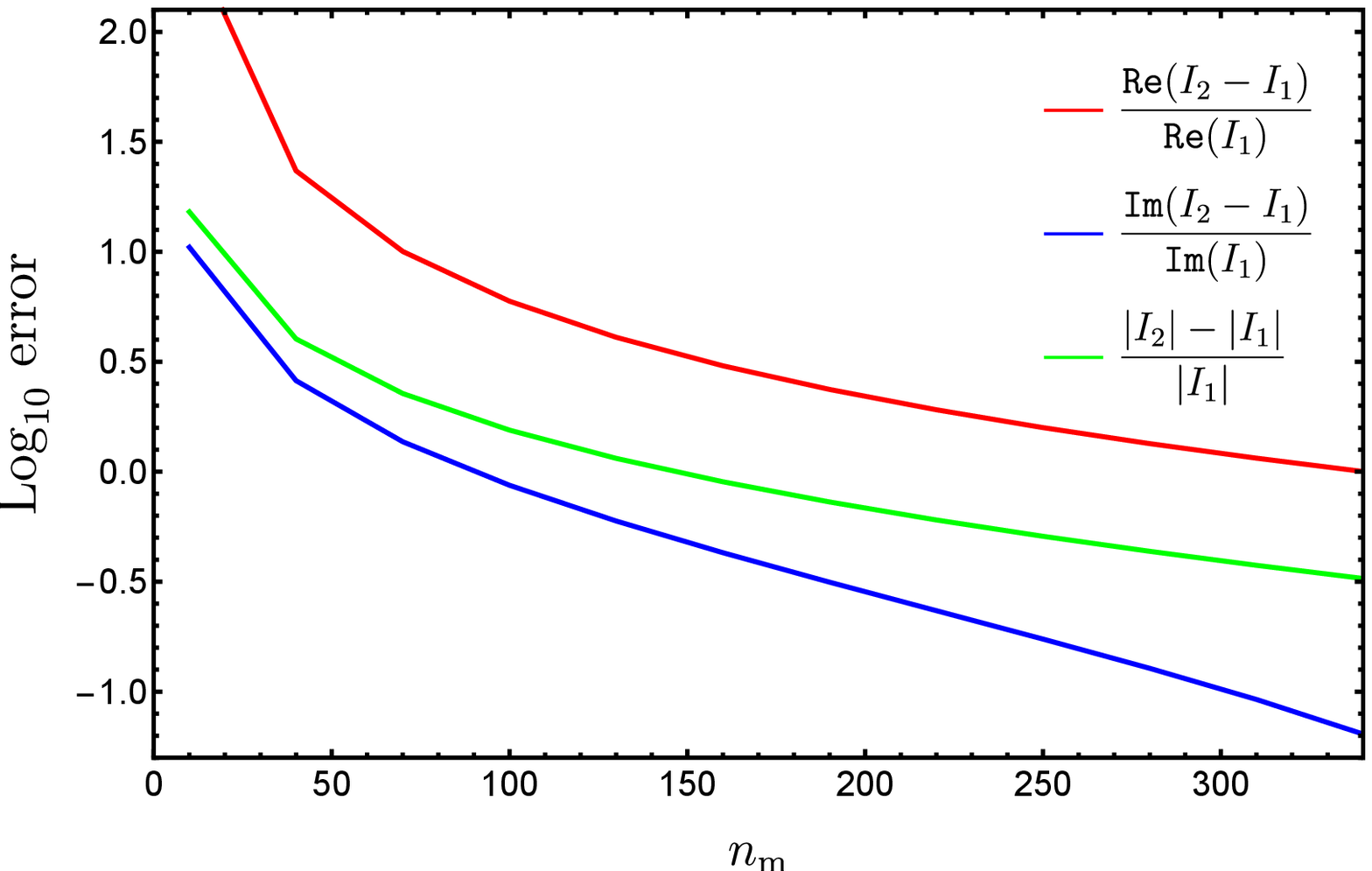}
\caption{
\label{fig:errorWnm}
Left: the dependence of the normalization error, $N_r$ of
Equation~\eqref{eq:Nrerror}, on the value of Hermite index $n$.  Right: the
dependence of the relative error, i.e., $(I_2-I_1)/I_1$, on the value of $n_m$
with $\mathtt{Re}(b) = \sqrt{ 2n_m +1}$ and $\mathtt{Im}(b) = \eta^{2/5}
\mathtt{Re}(b)$, with $\eta = 10^{-5}$.  The relative error in the approximate
integral is a complex function. Thus, we compare the error in the real part
(red), the imaginary part (blue) and the absolute value (green).  }
\end{figure}

Therefore, the error in the normalization due to our approximation is given by
\begin{eqnarray}
\label{eq:Nrerror}
N_r 
\equiv 
\frac{ \tilde{N} - N }{N}
=
\frac{e^{-n} n^n \sqrt{ \pi (2  n+1) }}{n!}-1.
\end{eqnarray}
On the left-hand side of Figure~\ref{fig:errorWnm}, we plot the error $N_r$ as a
function of the Hermite index $n$. It shows that our approximation produces a
relative error of less that 1\% for $n>20$.


However, when we compute the dispersion relation, the largest error in the integral comes from the difference between the analytical and approximate forms near the poles, i.e., near the solutions.
Therefore, we compare the values of the integral and its approximation near the expected solution.
The integral in Equation~[\ref{eq:fullDisp}] is
\begin{eqnarray}
I_1(b,n)
&=&
\frac{1}{ \int dy  H^2_n\left(y\right)   e^{-y^2 }  }
\int dy  \frac{ H^2_n\left(y\right)   e^{-y^2 } }{ ( y - b)^2   }
=
\frac{1}{\sqrt{\pi} 2^n n!}
\int_{-\infty}^{\infty} dy  \frac{ H^2_n\left(y\right)   e^{-y^2 } }{ ( y - b)^2   }.
\end{eqnarray}
This is approximated by 
\begin{eqnarray}
I_2(b,n)
&=&
\frac{
\int_{-\sqrt{2n+1}}^{\sqrt{2n+1}} dy    \left[ 1- \frac{y^2}{2n+1} \right]^{-\frac{1}{2}}  (y-b)^{-2} 
}
{
\int_{-\sqrt{2n+1}}^{\sqrt{2n+1}}  dy \left[ 
 1- \frac{y^2}{2n+1}
\right]^{-\frac{1}{2}} 
}
=
\frac{1}{2n+1}
\frac{
\int_{-1}^{1}  dz  \frac{  \left[ 1- z^2\right]^{-\frac{1}{2}} }{ ( z-b/\sqrt{2n+1})^2   }  
}
{
\int_{-1}^{1}  dz \left[ 
 1- z^2
\right]^{-\frac{1}{2}} 
}
\nonumber \\
&=&
\frac{b}{ \left( b^2- (2 n+1) \right)^{3/2}},
~~~~~~~~~~~~~~~~~~~
\end{eqnarray}
where we assumed that  $\mathtt{Im} (b) \neq 0 $.
Before comparing the values of the two functions, $I_1$ and $I_2$, for different
values of $n$, we first need to compute the characteristic value of their
complex and dimensionless argument, $b$, that enables a meaningful comparison. Using
\begin{eqnarray}
b^2 
&\equiv& 
\frac{\omega^2}{a^2 v_b^2}
=
\frac{\omega^2}{\omega^2_0}
\frac{\omega_0^2}{v_b^2} \sqrt{\frac{3}{\epsilon}} \frac{\sigma}{\omega_0}
=
\frac{\omega^2}{\omega^2_0}
\frac{\sigma^2}{v_b^2} \sqrt{\frac{3}{\epsilon  \lambda_D}} 
\sim
\frac{\omega^2}{\omega^2_0}  (2 n_m +1 ) 
\nonumber \\
\Rightarrow 
&&
\mathtt{Re}(b)  \sim \sqrt{ 2n_m +1}
~~ \& ~
\mathtt{Im}(b)  \sim \eta^{2/5} \mathtt{Re}(b) .
\end{eqnarray}

We show in the right panel of Figure~\ref{fig:errorWnm} the dependence of the relative error, i.e., $(I_2-I_1)/I_1$, on the value of $n_m$ for $\eta = 10^{-5}$.
The right panel of Figure~\ref{fig:errorWnm} shows that the error in the
approximation of the integral decreases as the value of $n_m$ increases.
The error in the imaginary part of the integral (which dictates the value of the
growth rates) is the smallest error and decreases exponentially fast.
This establishes the validity of our approximation of the integral to compute the fastest growth
rate in the large-$n$ limit.

\bibliography{refs}
\bibliographystyle{jpp}

\end{document}